\documentclass[preprint,12pt]{elsarticle}
\usepackage[margin=1in]{geometry}

\makeatletter
\def\ps@pprintTitle{%
  \let\@oddhead\@empty
  \let\@evenhead\@empty
  \def\@oddfoot{\hfill{\footnotesize\textit{\today}}}   % date
  \let\@evenfoot\@oddfoot}
\makeatother

\usepackage{xcolor}
\usepackage{color}
\usepackage{amsmath}
\usepackage{gensymb}
\usepackage{amssymb}
\usepackage{lineno}

\begin{document}

\begin{frontmatter}

\title{Mechanics of three-dimensional micro-architected interpenetrating phase composites}

\author[inst1]{Andrew Y. Chen}
\affiliation[inst1]{organization={Department of Mechanical Engineering},%Department and Organization
            addressline={Massachusetts Institute of Technology}, 
            city={Cambridge},
            postcode={02139}, 
            state={MA},
            country={USA}}

\author[inst1]{Carlos M. Portela\corref{cor1}}
\ead{cportela@mit.edu}
\cortext[cor1]{Corresponding author}

\begin{abstract}
\noindent Composite materials are used widely across engineering applications for their superior mechanical performance, a result of efficient load transfer between the structure and matrix phases. However, the inherently two-dimensional structure of laminated composites reduces their robustness to shear and out-of-plane loads, while unpredictable interlaminar failure and fiber pull-out can cause a sudden, catastrophic loss of load capacity. Meanwhile, recent advances toward uncovering structure-property relations in architected materials have led to structures with highly tunable mechanical properties, deformation, and even failure. Some of these architected materials have reached near-theoretical limits; however, the majority of current work focuses on describing the response of a single-material network in air, and the effect of adding a load-bearing second phase to a three-dimensional architecture is not well understood. Here, we develop facile fabrication methods for realizing centimeter-scale polymer- and carbon-based architected interpenetrating phase composite (IPC) materials, i.e., two-phase materials consisting of a continuous 3D architecture surrounded by a load-bearing matrix across length scales, and determine the effect of geometry and constituent material properties on the mechanics of these architected IPCs. Using these experiments together with computational models, we show that the matrix phase distributes stress effectively, resulting in a high-strength, stable response to loading. Notably, failure delocalization enhances energy dissipation of the composite, achieving specific energy absorption (SEA) values comparable to those of wound fiber tubes. Finally, we demonstrate that the stress state in an IPC can be tuned using geometric design and introduce an example of optimized mechanical response in an architected composite. Altogether, this work bridges the gap between mechanically efficient composites and tunable architected materials, laying the foundation for a new class of strong, resilient, and programmable materials.
\end{abstract}
\end{frontmatter}

% main text
\section{Introduction}
\label{sec:introduction}
Over the last two decades, composite materials have emerged as one of the preferred paradigms for robust and efficient structural materials, owing to their remarkable stiffness-to-weight and strength-to-weight properties. The majority of composites in service today are laminated structures consisting of bonded plies, each made up of long, oriented fibers held in place by a matrix material, commonly epoxy resin \cite{GARCIA2008}. These laminated sheets can be used alone or in a sandwich structure together with an extruded core; for example, honeycomb sandwich panels are prevalent in the automotive and aerospace industries. In each case, the structural elements (e.g., fibers) are typically made of a high-strength material such as carbon, glass, or aramid; this primary (reinforcing) phase is designed to withstand the applied loading on the composite. The secondary (matrix) phase, in comparison, is designed to bind the structural elements together and achieve load transfer between adjacent elements of the primary structure. This cooperative effect between material phases is the origin of the composite's extraordinary structural properties, such as energy absorption in the composite that is greater than the sum of each individual phase \cite{DMello2013, Han2017}.

Despite these advantages, composite materials suffer from several flaws owing to their layer-wise structure. Due to the nature of their oriented fibers or lay-up arrangement, laminated composites are inherently anisotropic, and even composites having a complex ply structure remain weak in shear or out-of-plane loading compared to in-plane tension. Moreover, fiber composites often fail by fiber pull-out or delamination between plies \cite{Wisnom2009}, and similarly sandwich panels are prone to buckling, cohesive fracture in the sandwich layer, or debonding at the interface \cite{Xiong2016}. These failure events tend to be catastrophic, associated with a sudden drop in load capacity of the structure with minimal warning in the form of inelastic deformation. Moreover, due to the presence of defects in manufacturing or processing, failure can happen at applied stresses much lower than the theoretical strengths of the composites. Taken together, these conditions require that care is taken in the design and fabrication of composite materials to avoid catastrophic failure, especially in safety-critical components.

Meanwhile, developments in the field of architected materials have yielded a thorough understanding of the structure-property relations in periodic architected materials, i.e., multiscale materials consisting of a unit cell tessellated in three-dimensional space \cite{Fleck2010}. Such architected materials have been found to exhibit near-ideal specific stiffness \cite{CrookEtAl2020}, high strength and toughness \cite{BauerEtAl2016, GuellIzard2019}, and other favorable material properties. Importantly, the mechanical behavior of such architected materials is highly tunable through control of the unit cell's morphology. One mechanism to improve their performance is by scaling down to take advantage of so-called ``size effects'', whereby a reduction in the feature size (e.g., to the sub-micron scale) yields an enhancement in properties or in mechanical robustness  \cite{MezaEtAl2014,Bauer2017}. However,  to measure a \textit{material} response (cf. a \textit{structural} response in a finite specimen), adequate separation of scales must be attained in the architected material. Recent advances in small-scale fabrication technology, e.g. using projection microstereolithography (PuSL) or two-photon lithography (TPL) processes,  have enabled the realization of architected materials with such a separation of scales \cite{Zheng2016, Shaikeea2022, Serles2025}.

In recent years, examples of architected composite materials have emerged which address the extreme anisotropy and catastrophic failure modes of traditional laminated composites, introducing instead a high degree of tunability in the performance of the composite until failure. Although the concept of a woven-fiber composite has existed for decades \cite{Scida1999}, one early example of a three-dimensional architected composite material is the so-called ``non-woven composite'' \cite{Meza2019}. These composites exhibited stable behavior under compression and a high degree of compressive ductility, despite the formation of localized strain in the form of kink bands, proving the applicability of the load transfer concept beyond laminate constructions. Later work on three-dimensional composite materials led to the widespread use of additive manufacturing (AM), among other techniques, to create the scaffold structure of a so-called ``interpenetrating phase composite'' (IPC), in which two unique material phases (which we will call the ``reinforcement'' and ``matrix'') each form their own self-connected network  \cite{Wegner2000}. Although the structure and properties of an IPC have been studied for decades, the availability of commercial AM solutions in the past decade has enabled the development of a corpus of work which explicitly studies the effect of architecture-phase morphology on mechanical properties. Specifically, in recent years, beam-based periodic, surface-based periodic, and aperiodic IPCs have been studied \cite{Seetoh2019, zhang2021, Li2018, AlKetan2017, Carlsson2024_exp, Wang2024, Singh2023, Raj2024, Fox2024, Singh2024}, demonstrating that the framework of architected interpenetrating phase composites can produce materials with not only desirable but also tunable mechanical properties. Namely, it has been shown that architecture may be harnessed to optimize stiffness, strength, toughness, and damage resistance, yielding composites that have superior properties to monolithic architected or natural materials.

Nevertheless, there remain two important limitations on the existing literature on interpenetrating phase composites. First, most studies on architected IPCs have considered a structure where the architecture phase is made of a polymer. This is consistent with the use of commercially available 3D-printing systems and the facility with which freeform polymer samples can be fabricated. Although polymeric materials compatible with AM systems span a range of mechanical properties, they are generally orders of magnitude less stiff and strong compared to the principal structural materials from which traditional fiber and sandwich composites are made (viz. carbon, glass, or aramid). Consequently, there is a lack of direct comparison between state-of-the-art architected IPCs and traditional structural composites, hindering the possible advancement of this novel technology. 

Moreover, due in part to feature-size limitations on AM systems combined with scale limitations on commercial mechanical testing frames, the majority of current examples of IPCs fail to attain a separation of scales. In turn, the deformation and failure behavior observed during empirical testing is strongly influenced by the proportion of unit cells experiencing boundary effects (e.g., due to free boundary conditions at the lateral surfaces or friction at the compression platens). As a result, the mechanical properties reported as a result of this \emph{finite-tessellation} testing often represent structural properties rather than material properties.

In this work, we directly address these shortcomings by introducing a facile, material-agnostic fabrication system for multi-scale architected interpenetrating composites and experimentally and computationally investigate the influence of constituent material properties on their mechanical behavior. We demonstrate that architected IPCs display a stable response to applied loading, giving rise to a high specific energy absorption capacity, comparable with that of structural wound fiber tubes. We demonstrate fabrication of not only polymer-based architectures, for benchmarking and for a direct comparison to the existing literature, but also carbon-based architectures, to broaden the space of micro-architected composite materials and for a direct comparison to traditional laminated composites. In each case, we achieve a separation of scales while retaining a centimeter-scale specimen size, enabling the measurement of material properties. To develop a fundamental understanding of the mechanics that underlies their behavior, we experimentally and computationally characterize the performance of these composites by contextualizing the elastic properties of the composite as a function of the ratio of elastic moduli between the phases. Through this analysis, we demonstrate that a critical stiffness fraction exists beyond which stretching-dominated behavior is observed in an otherwise kinematically non-rigid reinforcing 3D architecture. Next, we investigate the nonlinear behavior of the composites, demonstrating how the progressive, three-dimensional nature of their deformation leads to a stable, tough failure mechanism. Finally, we show an example of tunable energy absorption in an architected composite by exploiting the incompressible nature of a candidate matrix material together with an auxetic reinforcing phase. Altogether, this fundamental study of the complex interplay between morphology, material selection, and mechanics lays the foundation for highly tunable and scalable IPCs beyond polymer systems toward a myriad of applications.

\section{Fabrication of architected IPCs}
\label{sec:materials}
\begin{figure}[h!]
    \centering
    \includegraphics[width=0.95\linewidth]{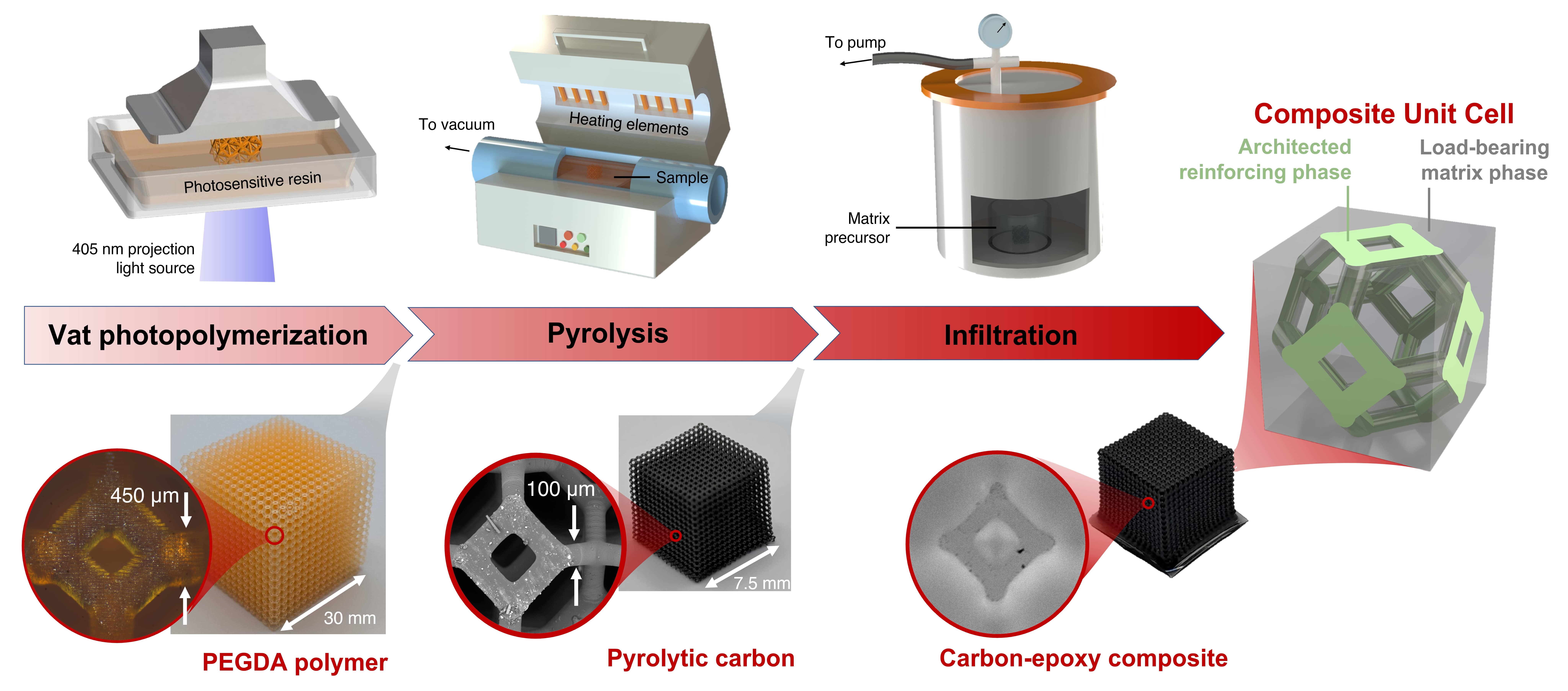}
    \caption{Overview of carbon-epoxy micro-architected interpenetrating phase composite fabrication by vat photopolymerization, pyrolysis, and subsequent matrix infiltration.}
    \label{fig:fabrication}
\end{figure}

To achieve centimeter-scale architected interpenetrating phase composites with a proper separation of scales, we first fabricated the 3D architected reinforcing phase and then performed a matrix-infiltration step to create an architected composite (Figure \ref{fig:fabrication}). This sequential fabrication method is compatible with a wide variety of materials and structures, expanding the achievable property and morphology space. We focused on two material systems: one having a polymer reinforcing phase, printed using two-photon lithography (TPL); and separately, one having a carbon reinforcing phase, created using vat photopolymerization (VP) 3D-printing followed by pyrolysis. 

\paragraph{Polymer reinforcing phase} We used a TPL printer NanoOne (UpNano, Vienna, Austria) to fabricate microscale $20 \times 20 \times 20$ tessellations of unit cells out of an acrylate-based two-photon photosensitive resin UpPhoto. The unit cells were designed in a computer-aided design (CAD) software to range in relative density from 7\% to 22\%. Each unit cell nominally measured 200 \textmu{}m in length, and hence the nominal specimen size was 4 mm. However, fabricated specimens were of slightly higher relative density compared to these nominal values (viz., 10\% to 28\%) due to the printing process.

\paragraph{Carbon reinforcing phase} We used a desktop ultraviolet liquid crystal display (UV LCD) 3D-printer (Phrozen Tech Co., Ltd., Hsinchu City, Taiwan) to fabricate macroscale $15 \times 15 \times 15$ tesselations of unit cells out of a custom, poly(ethylene glycol) diacrylate (PEGDA)-based photopolymer resin. These unit cells were designed in CAD to range in relative density from 4\% to 12\% with the understanding that the relative density post-pyrolysis would increase. Unlike the TPL structures, which varied only in beam diameter as the relative density was changed, these structures were designed to be of constant mass for all relative densities to ensure consistent and uniform pyrolysis. To this end, both beam diameter and unit cell size were varied as relative density changed; beam thicknesses ranged between 230 and 310 \textmu{}m, and unit cell sizes ranged between 1800 and 2850 \textmu{}m. After printing, washing, and drying, the PEDGA structures were transferred to a high-temperature vacuum furnace for pyrolysis. The chamber was evacuated to a pressure not exceeding 100 mTorr, and the temperature followed a prescribed heating profile which was constructed following thermogravimetric analysis (TGA) on the PEGDA-based resin (Section \ref{sec:pyrolysis}). It has been shown that pyrolytic carbon undergoes microstructural changes with a corresponding change in Young's modulus up to a temperature of 900$\degree$C, with stablization (in both microstructure and modulus) above this temperature \cite{Sharma2018}. For this reason, our heating cycle includes a hold at a maximum temperature of 1000$\degree$C. During pyrolysis, the polymer samples experienced shrinkage, corresponding to an average mass loss of 98\% and average linear shrinkage by a factor of 4.5, yielding centimeter-scale carbonized specimens having a characteristic beam diameter of $\sim$100 \textmu{}m.

After fabrication of the polymer- or carbon-based reinforcing phase, we created composites using a matrix infiltration process. We used three matrix materials: urethane resin Smooth-Cast 61D (SC61D), silicone elastomer polydimethylsiloxane (PDMS), and epoxy resin West System 105/206. Each of these materials is commercially available as a two-part liquid precursor containing a base and a crosslinker; solidifcation occurs after a short time when the two parts are mixed. Matrix infiltration was accomplished using vacuum infiltration at approximately 5 Torr (for the PDMS and SC61D matrix materials) or by centrifuge infiltration at 2000 RPM (for the epoxy resin, which was too viscous in the liquid state to be compatible with vacuum infiltration). In both cases, we submerged the printed or carbonized specimens in approximately 10 mL of matrix precursor directly after mixing the two parts. Evacuation or centrifuging continued until the pot life of the matrix material specified by the manufacturer was nearly reached. The infiltrated composites were then extracted from the liquid, excess material was gently wiped from the surface, and samples were then cured according to the manufacturer's instructions. Following complete curing, each sample was checked for voids using X-ray computed tomography (XCT). All composites samples used for mechanical characterization had a volumetric void fraction not exceeding 1\% (Section \ref{sec:voids}).

\section{Linear-elastic behavior}
\subsection{Scaling laws}
\label{sec:elasticprops}
Seminal works on the response of beam-based architected materials have produced well-established results on their mechanical behavior as a function of morphology \cite{DeshpandeEtAl2001a, Fleck2004, Ashby2006}. In particular, for a unit cell made of slender beams, the relative density (i.e., fill fraction) $\tilde{\rho}$ is related to the relative stiffness (i.e., the nondimensional ratio of the effective stiffness of the unit cell to the stiffness of the constituent material) $\tilde{E}$ by 

\begin{equation}\label{eq:scalinglaw}
    \tilde{E} = C\tilde{\rho}^m,
\end{equation}

where $C$ is a dimensionless fitting constant. The exponent $m$ varies depending on whether beams in the unit cell tend to undergo a \textit{stretching}- or \textit{bending}-dominated response; in particular, $m = 1$ for ideal stretching-dominated structures, and $m = 2$ for ideal bending-dominated structures \cite{GibsonAshby1999}---leading to the classical stiffness scaling laws for beam-based architected materials. 
Therefore, maximization of the structural efficiency in terms of linear-elastic stiffness is often sought by designing stretching-dominated architectures. In reality, however, fabricated specimens behave in a non-ideal manner owing to finite slenderness (which undermines Euler-Bernoulli-beam assumptions that lead to the classical scaling laws), and because the geometry of intersecting beams (i.e., the shape of nodes in the lattice) result in the propagation of bending moments. Consequently, the scaling exponent for most manufacturable architected materials can differ from the theoretical values, and are typically non-integers \cite{PortelaEtAl2018,MezaEtAl2017}. Nevertheless, an analysis of the stiffness-density scaling behavior in a beam-based architected material offers important insight into its structural behavior.

A necessary condition for a given beam-based geometry to exhibit stretching-dominated behavior is the satisfaction of Maxwell's criterion, $b - 3j + 6 \geq 0$ in three dimensions, where $b$ is the number of beams and $j$ is the number of nodes in a unit cell \cite{Ma2022, PellegrinoCalladine1986}. The sufficient condition requires that the structure admit no zero-energy mechanisms if pin-jointed. Hence, a structure that fails to satisfy Maxwell's criterion must be bending-dominated along a particular loading direction. The Kelvin cell (i.e., regular tetrakaidecahedron \cite{Weaire2009}), which we consider in this work, is such a structure that has served in the literature as an example of a bending-dominated architecture \cite{MezaEtAl2017, Chen2023}. Moreover, throughout this work we consider tessellations of Kelvin cells where the number of unit cells well exceeds the experimentally- and numerically-determined convergent value for the Kelvin cell \cite{Wang2023_2}, guaranteeing a homogenized material response.

\subsection{Effect of the composite stiffness fraction}
\begin{figure}
    \centering
    \includegraphics[width=0.9\linewidth]{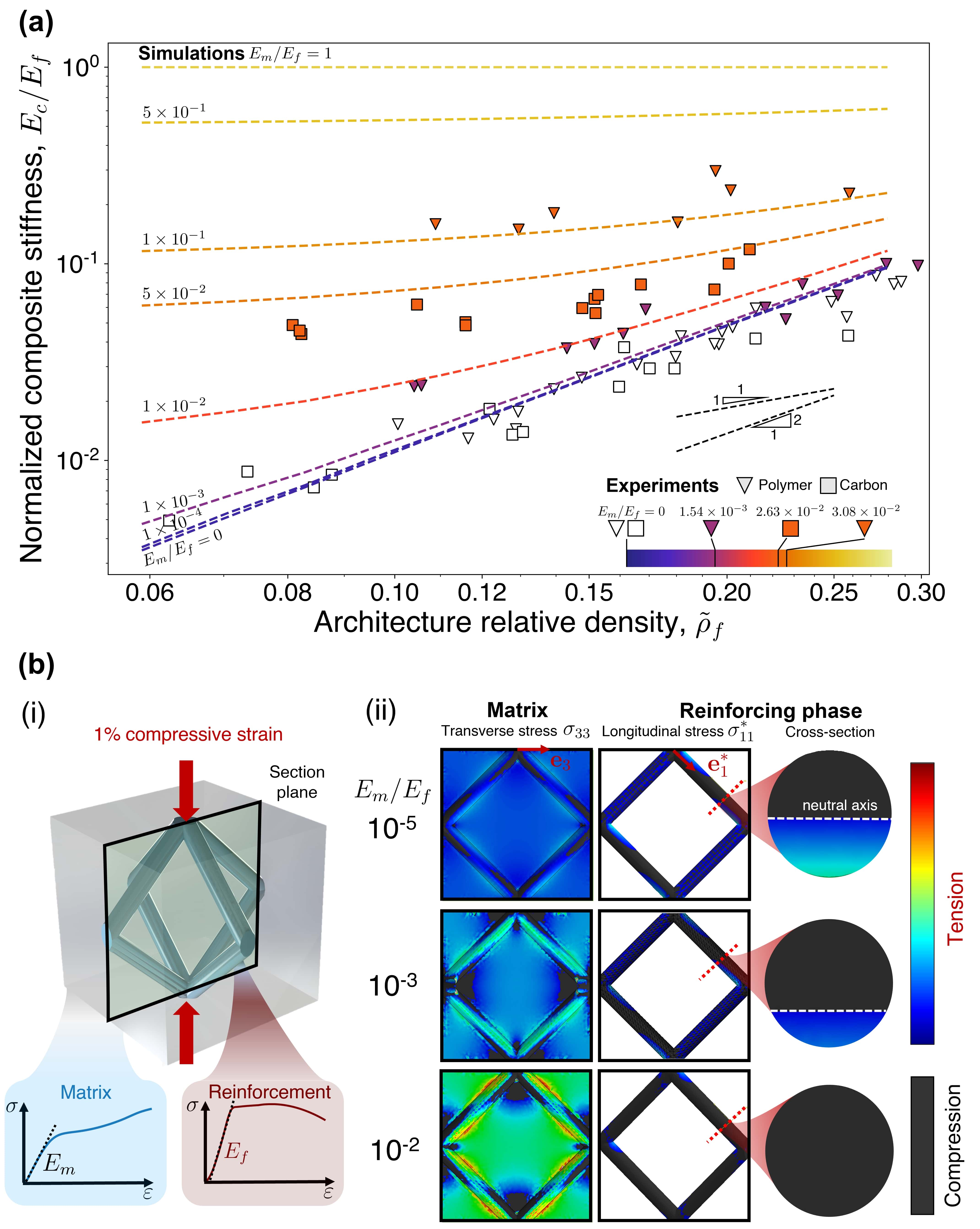}
    \caption{Elastic behavior of architected IPCs. (a) Plotting the normalized effective composite stiffness $E_c/E_f$ as a function of reinforcing phase relative density $\tilde{\rho}_f$ for varying reinforcement-matrix stiffness fractions $E_m / E_f$ suggests a transition from a bending-dominated behavior to a stretching-dominated behavior in a non-rigid architecture. (b) Understanding this transition using a simplified unit-cell model (i), which demonstrates that as $E_m / E_f$ is increases towards the critical value, the bending state is gradually replaced by a stretching state as a transverse path of load transfer is established in the matrix (ii).}
    \label{fig:elastic}
\end{figure}

To quantify the effect of adding a load-bearing matrix phase to a material consisting of tessellated Kelvin cells, we parameterize the Young's modulus of the architecture phase as $E_f$ and the Young's modulus of the matrix phase as $E_m$, with $E_m \leq E_f$ in accordance with the standard assumption for composite materials. 
% We assume, as is typical for two-phase composites, that $E_m \leq E_f$. 
Denoting the effective stiffness of the resulting composite as $E_c$, and simulating the response of a Kelvin cell to a linear perturbation at 1\% uniaxial compressive strain, Fig.~\ref{fig:elastic}(a) plots the nondimensional value $E_c/E_f$ as a function of the architecture-phase relative density $\tilde{\rho}$ for eight values of $E_m / E_f$, ranging from $E_m / E_f = 0$ (equivalent to the classical single-material case) to $E_m / E_f = 1$ (i.e., the case where the material is elastically homogeneous).

As shown in Fig~\ref{fig:elastic}(a), the effective stiffness of the composite increases with relative density, as predicted by the classical arguments, and increases with increasing $E_m / E_f$, as predicted by a rule-of-mixtures approach. Moreover, it is bounded from below by the single-material case, where we recover an approximately quadratic scaling law for the Kelvin cell (here, $m = 2.182$ in the slender limit, in good agreement with previous work \cite{MezaEtAl2017, Arretche2018}), and bounded from above by the case $E_m / E_f = 1$, where trivially $E_c = E_m = E_f$ and the effective scaling exponent is zero.

Therefore, this indicates that there exists a critical stiffness fraction $0 < E_m / E_f < 1$, which we denote $(E_m / E_f)^*$, for which $m \approx 1$. Table \ref{tab:tab1}, which contains the best-fit scaling exponents in the slender regime (denoted $m$ and identified to be the first region of constant slope for $\tilde{\rho} \leq 12\%$), identifies that for $E_m / E_f \geq 1 \times 10^{-2}$, the composite Kelvin cell has a scaling behavior described by $m \leq 1$. Table \ref{tab:tab1} also contains the best-fit exponent in the non-slender regime (denoted $\hat{m}$), which coincides with the range of relative densities where nodal geometries are non-negligible and where the material behavior cannot be assumed to be in purely bending- or stretching-dominated modes.

\begin{table}[h!]
	\centering
	\begin{tabular}{|c|c|c|} \hline
		\begin{tabular}[c]{@{}c@{}}Stiffness fraction\\ $E_m / E_f$\end{tabular} & \begin{tabular}[c]{@{}c@{}}Scaling exponent $m$\\ $(\tilde{\rho} \leq 12\%)$\end{tabular} & \begin{tabular}[c]{@{}c@{}}Exponent $\hat{m}$\\ $(\tilde{\rho} > 12\%)$\end{tabular} \\ \hline \hline
		0 & 2.182 & 2.084 \\
		$1 \times 10^{-4}$ & 2.140 & 2.083 \\
		$1 \times 10^{-3}$ & 1.885 & 2.020 \\
		$1 \times 10^{-2}$ & 0.948 & 1.615 \\
		$5 \times 10^{-2}$ & 0.377 & 0.907 \\
		$1 \times 10^{-1}$ & 0.248 & 0.606 \\
		$5 \times 10^{-1}$ & 0.063 & 0.140 \\
            $1$ & 0 & 0 \\ \hline
	\end{tabular}
	\caption{Computed scaling-law exponents $m$ for the composite Kelvin cell as a function of stiffness fraction $E_m / E_f$. In each case, the exponent changes above a critical relative density of 12\%, where the slender-beam assumption fails to be satisfied; this secondary exponent is denoted $\hat{m}$.}
	\label{tab:tab1}
\end{table}

While the classical slender-beam theory identifies linear scaling for the case where all beams within a unit cell exhibit stretching, this linear-elastic analysis indicates that a Kelvin cell---which fails to satisfy Maxwell's criterion in the single-material case---exhibits linear scaling in composites of $E_m / E_f \geq 10^{-2}$. To investigate this phenomenon and determine whether the converse is true (i.e., whether $m = 1$ in the scaling behavior implies stretching-dominated beams), we investigated a simpler three-dimensional model inspired by the two-dimensional ``minimal example'' of Deshpande, Ashby, and Fleck \cite{DeshpandeEtAl2001a}, depicted in Fig.~\ref{fig:elastic}(b)(i). Our model consists of a non-rigid structure of maximal aspect ratio such that the slender-beam condition is well approximated. In particular, if our model were pin-jointed, a zero-energy mechanism would exist by which the lateral nodes can displace outward as the unit cell is placed in uniaxial compression.

Using this simplified geometry, we performed the same linear perturbations as on the Kelvin cell, with identical boundary conditions and for the same range of $E_m / E_f$ values. To determine the state of stress in the beams after the deformation is applied, we make a cut along the beam cross-section and define a local coordinate system having longitudinal axis $\mathbf{e}_1^*$ oriented along the beam. With respect to this coordinate system, observation of $\sigma_{11}^*$ across the beam cross-section is sufficient to determine whether the beam is in a state of bending or stretching. Namely, the presence of a neutral axis whereby one portion of the cross-section has $\sigma_{11}^* < 0$, and another portion has $\sigma_{11}^* > 0$, indicates bending (Fig.~\ref{fig:elastic}(b)(ii)).

When $E_m / E_f \ll (E_m / E_f)^*$, a neutral axis is present, coinciding with the geometric centroid of the beam---as expected for a beam in pure bending (i.e., $E_m / E_f \rightarrow 0$). When $E_m / E_f < (E_m / E_f)^*$, a neutral axis is present but is shifted relative to the centroid, indicating a superposed bending-stretching state. However, for $E_m / E_f \geq (E_m / E_f)^*$, the neutral axis vanishes, and the entire cross-section experiences a state of compression.

Correspondingly, observing the transverse state of stress in the matrix material offers an explanation for this transition. In the architecture, the addition of horizontally-oriented beams through the center of the specimen would remove the aforementioned mechanism and render the structure stretching-dominated \cite{DeshpandeEtAl2001a}. As the stiffness fraction $E_m / E_f$ is increased to the critical value, it is clear that the transverse stress taken up by the matrix increases accordingly, until a pathway for stress across the center of the unit cell develops---exactly where the horizontal beam would have existed. This evidence indicates that a matrix of sufficient stiffness relative to the reinforcement's constituent stiffness aids the establishment of a load transfer path which acts like a ``missing'' beam in the reinforcement, and explains its transition to a state of stretching.

\subsection{Experimental validation}
To verify the existence of a critical stiffness fraction by which the bending-to-stretching transition is accomplished, we extracted stiffness metrics from the uniaxial compression experiments on the fabricated specimens. The two sets of uninfiltrated specimens (viz., polymer and carbon architectures) represent the case of $E_m / E_f = 0$, whereas the three sets of composite specimens had stiffness fractions $E_m / E_f = 1.54 \times 10^{-3}$ (polymer/PDMS), $E_m / E_f = 2.63 \times 10^{-2}$ (carbon/epoxy), and $E_m / E_f = 3.08 \times 10^{-2}$ (polymer/urethane). In particular, we designed two sets of specimens having $E_m / E_f$ above the critical value of $10^{-2}$ from different constituent materials to demonstrate that the stiffness fraction phenomenon is effectively independent of nonlinear material behavior.

Experiments on uninfiltrated specimens corroborate the computational results; in particular, they confirm that the architecture exhibits a bending-dominated scaling response, independent of the constituent material ($m = 1.875$ for the polymer lattices and $m = 1.725$ for the carbon lattices). Moreover, the three sets of composite lattices demonstrate the expected stratification based on the stiffness fraction, as predicted by the simulations (Fig.~\ref{fig:elastic}(a)). The polymer/PDMS system, having a sub-critical stiffness fraction, has a best-fit exponent of 1.307, indicating that fully stretching-dominated behavior is not achieved. On the other hand, both systems with $E_m / E_f > 10^{-2}$ have scaling exponents less than unity: 0.672 for the carbon/epoxy system and 0.602 for the polymer/urethane system. This suggests that the hypothesized stress transfer pathway does exist in these composites of sufficient matrix stiffness, resulting in a shift from bending-dominated behavior toward stretching-dominated behavior.

It must be mentioned that most fabricated specimens had relative densities outside the slender-beam regime, so the best-fit exponents do not directly represent their corresponding scaling exponents in the low relative-density limit. Namely, they cannot directly imply a state of pure bending or pure stretching. However, as indicated by the computational results (Table \ref{tab:tab1}), there is a strong correlation between the slender-beam-limit scaling exponent and the best-fit exponent outside this regime. Thus, these experiments still serve as evidence that there exists a bending-to-stretching transition in the fabricated composites.

\section{Nonlinear behavior}
\label{sec:nonlinear}
We consider now the behavior of the architected IPCs past the elastic regime, where the stress-strain response turns markedly nonlinear. As the behavior of polymer-polymer IPCs has been studied extensively throughout the literature, we focus primarily on the behavior of the carbon-based specimens; the behavior of the polymer system is briefly described in the SI Appendix and generally agrees with observations made in other studies. We begin by benchmarking the performance of the composite using the behavior of the architected reinforcing phase alone.

\subsection{Behavior of the carbon reinforcing phase}
\begin{figure}
    \centering
    \includegraphics[width=0.9\linewidth]{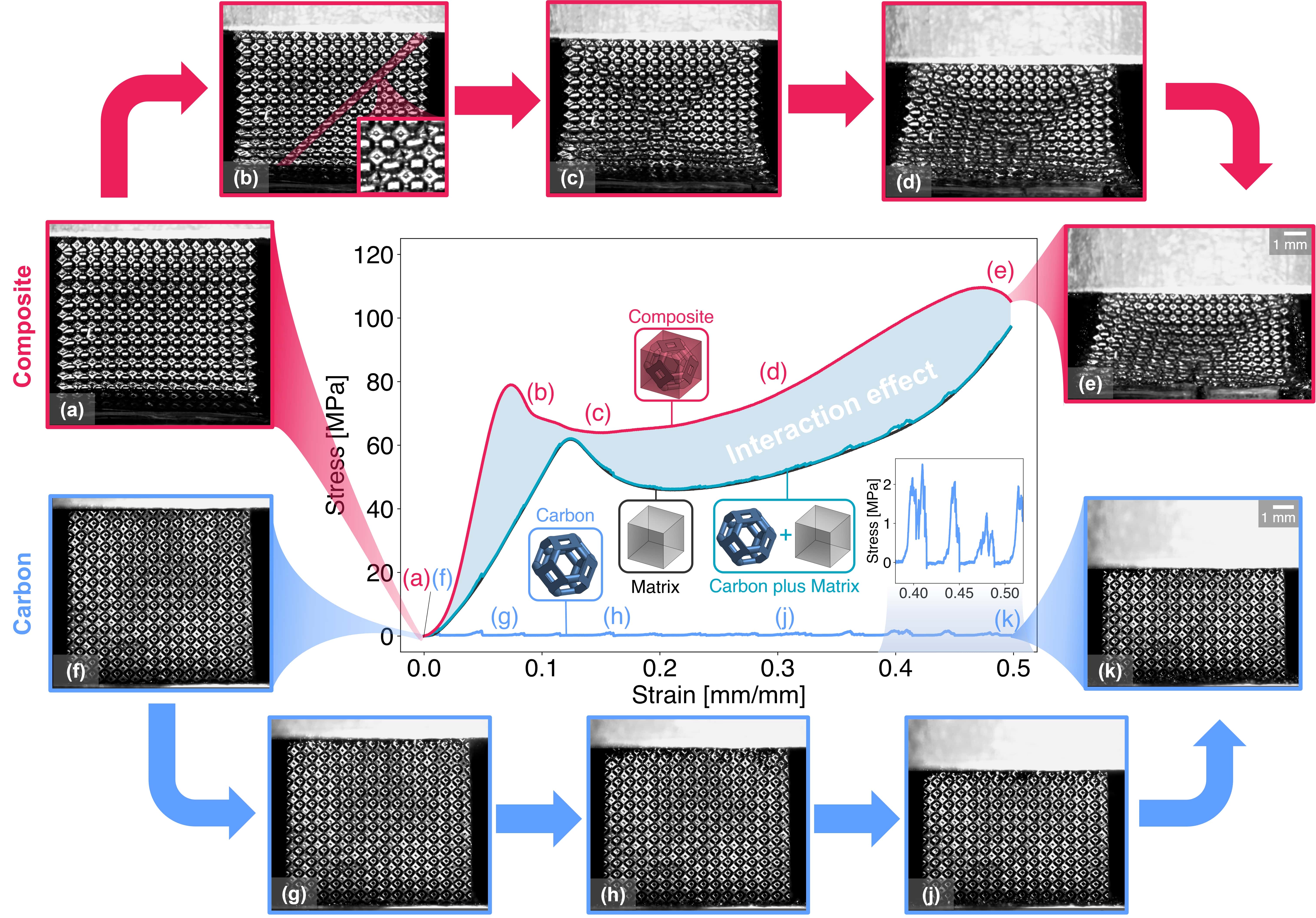}
    \caption{Representative mechanical behavior of the carbon-epoxy composite in uniaxial compression. Under increasing strain, (a)--(e), the composite demonstrates the emergence of double shear bands, followed by stable crushing and compaction. In comparison, the carbon reinforcing phase alone demonstrates hyperlocalized strut fracture, (f)--(k), contributing to an unstable, low-strength behavior. Also plotted is the stress-strain behavior of the monolithic matrix material, demonstrating that the composite's performance exceeds the sum of its constituent parts (the interaction effect).}
    \label{fig:stress_strain}
\end{figure}

The mechanical response of the carbon reinforcing phase alone is shown in Fig.~\ref{fig:stress_strain}(f)--(k) and is typical of all structures tested, independent of relative density (which ranged from 8\% to 22\% in the fabricated samples) or specimen size. In particular, experiments reveal that the method of failure of the carbon Kelvin cell is characterized by local failure dominated by layer-by-layer compaction. Since pyrolytic carbon fails by brittle fast fracture, the carbon Kelvin lattices effectively experience layer-wise fracture in a sequential manner from top to bottom, in a manner consistent with existing data on the compression of pyrolytic carbon architectures \cite{CrookEtAl2020, Wang2022, Zhang2019a}. This response likely emerges from friction imposed by the compression platen which locally increases the magnitude of the stress state at the top layer of unit cells, causing initiation of failure at those sites of highest stress concentration. Hence, unit cells appear to be crushed one layer at a time in a layer-by-layer fashion, while the remainder of the specimen outside of the crush band remains effectively pristine.

The sawtooth pattern in the measured stress-strain response reflects this behavior. Although the stress response appears approximately periodic, the apparent frequency does not match the number of unit-cell layers, and post-mortem characterization reveals that an entire layer of unit cells does not fracture all at once. Rather, once fracture is nucleated somewhere in the crush band, the extent of this ``hyperlocal'' failure determines how stress is transmitted to the next layer of unit cells and consequently the extent and nature of failure there. The high-frequency component of the stress-strain response, which appears to be superposed onto the quasi-periodic behavior, reflects the hyperlocality of this effect (namely, the fracture of individual beams).

Finally, it is seen that after the initial load peak is reached (corresponding to the complete failure of the top-most layer of unit cells), the peak value of stress is not attained at any subsequent deformation state. This suggests that during the failure of subsequent layers (e.g., Fig.~\ref{fig:stress_strain}(g)--(j)), there is never a point when an entire layer of pristine unit cells is simultaneously able to carry the load. Moreover, it implies that the peak load coincides with the critical crushing strain of a single layer of unit cells, and thus the load capacity of the entire structure is effectively determined only by this single layer. This behavior is mechanically inefficient, defect-sensitive, and unstable, but offers a multitude of opportunities for optimization.

\subsection{Behavior of the carbon-epoxy composite} The compression response of the carbon-epoxy composite Kelvin lattice, shown in Fig.~\ref{fig:stress_strain}(a)--(e), demonstrates a direct contrast to the brittle response of the reinforcing-phase alone (Fig.~\ref{fig:stress_strain}(f)--(k)). 
As the relative density of the reinforcing phase increases, the stiffness and strength of the composite increase (Section \ref{sec:rd}). Generally, the stress-strain response of the composite consists of an initial linear region associated with elastic deformation, followed by a first force peak corresponding to a local maximum in load. After a brief reduction in load, the stress passes through a second stable phase until a second peak around 50\% macroscopic strain is reached, after which the load drops in a manner consistent with ultimate failure. At low and intermediate reinforcing-phase relative densities ($\tilde{\rho}_f \lesssim 20\%$), this second stable phase consists of increasing stress with increasing strain; at the highest relative densities, the second phase is instead characterized by a load plateau. In comparison to the single-material case, the composite is capable of withstanding a significant amount of load before the onset of nonlinearity; the effective strength of the composite, taken to coincide with the first load peak, is typically 30 to 50 times greater than the peak stress attained by the reinforcement alone.

Optical imaging during uniaxial compression reveals that the layer-wise compaction mechanism displayed by the single-material carbon architecture is entirely suppressed. Rather, during the elastic region, the composite deforms largely homogeneously. The first load peak is associated with the development of a primary shear band across the front face of the specimen (Fig.~\ref{fig:stress_strain}(b)) characterized by damage local to unit cells contained within the band. 
Soon after the development of this primary shear band, a secondary shear band directed across the other diagonal appears in view, coinciding with the beginning of the second stable phase in the stress-strain curve (Fig.~\ref{fig:stress_strain}(c)). 
This double shear-banding behavior was observed among all such carbon-epoxy composites, regardless of relative density or specimen size; however, the orientation of the primary and secondary shear bands (i.e., which diagonal plane sheared first) was arbitrary. We show later (Section \ref{sec:xct}) that the secondary shear band visible in the optical image is only one of a family of secondary shear bands that form across diagonal planes of the specimen. 

Following the development of the secondary shear bands, the composite specimens transitioned into a crushing regime whereby unit cells below the shear bands appear to take up the majority of the damage (Fig.~\ref{fig:stress_strain}(d)), corresponding to an increase in the engineering stress taken up by the specimen (in the latter half of the second stable regime, where a measure of true stress is difficult to determine due to pervasive localized deformation). 

Finally, when the crushing behavior leads to crack formation (Fig.~\ref{fig:stress_strain}(e)), the stress reaches the second, final load peak before ultimately decreasing. This second load peak, which typically occurs near a macroscopic strain of 50\%, marks the failure strain of the composite specimen due to the initiation of macroscopic cracks.

\subsection{Failure of the carbon composite system}\label{sec:xct} 
\begin{figure}
    \centering
    \includegraphics[width=0.95\linewidth]{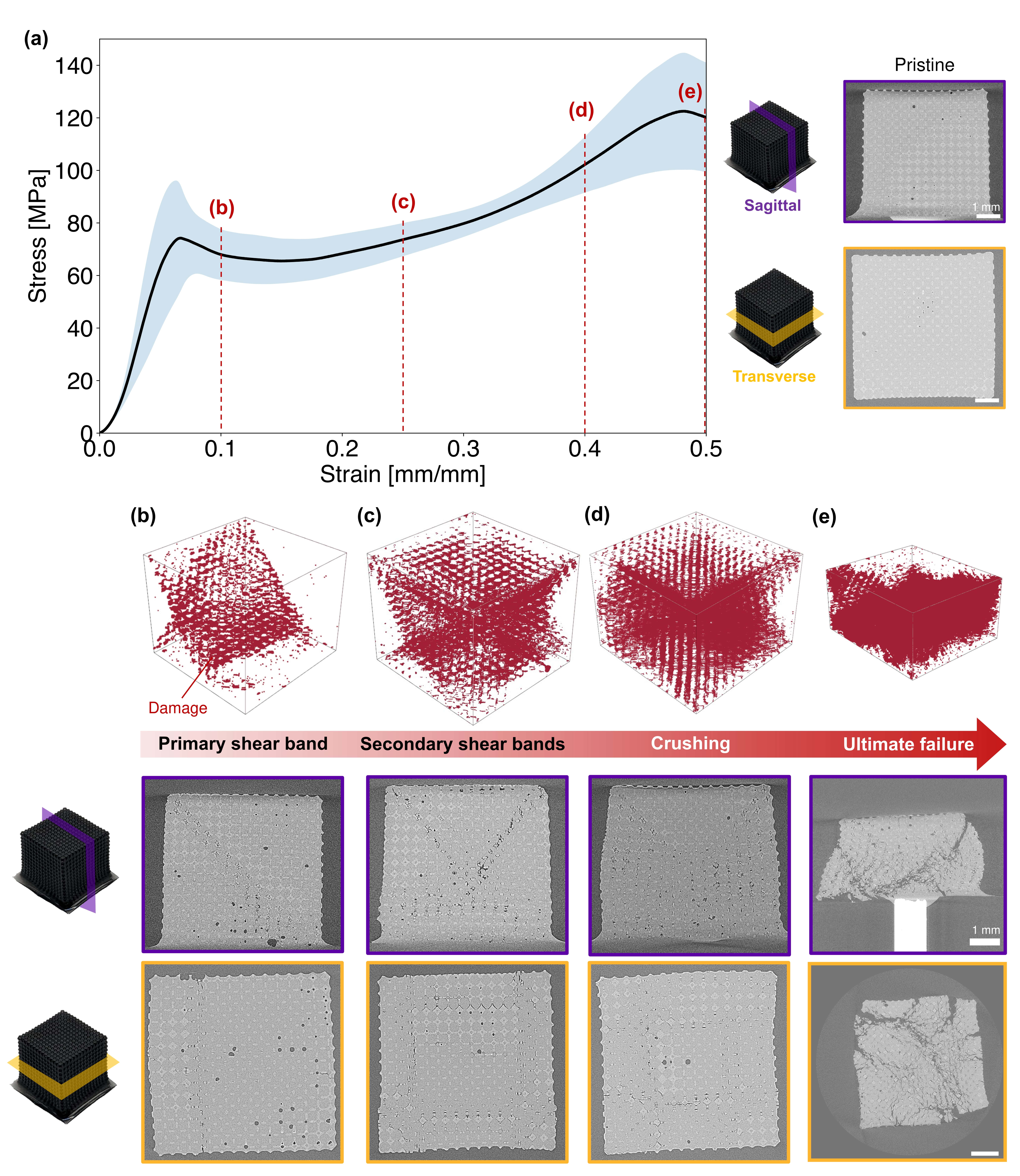}
    \caption{Sequential X-ray computed tomography of composite samples under increasing macroscopic strain, ranging from 0\% to 50\% macroscopic strain, at points shown on the aggregate stress-strain response (a). As deformation progresses, we observe the formation of an initial shear band at the first load peak (b), secondary shear bands (c), and crushing (d) followed by ultimate failure (e). Together with reconstructions in the transverse (yellow) and sagittal (purple) planes, three-dimensional visualizations reveal the location of internal cracks (red) in the composite body, both interfacial and cohesive (primarily within the carbon system).}
    \label{fig:XCT}
\end{figure}

To visualize the three-dimensional nature of the composite failure, as well as determine deformation beyond that of the specimen free boundaries, we performed X-ray computed tomography (XCT) scans on pristine (untested) composite samples, as well samples taken to several prescribed macroscopic strains. The chosen strains coincided with major events---namely the formation of shear bands---as indicated in Fig.~\ref{fig:XCT} suggested both by optical imaging and by features of the stress-strain response. To this end, Fig.~\ref{fig:XCT}(a) shows an ``aggregate'' stress-strain response, computed by averaging the responses of samples having varying reinforcement-phase relative densities, and with a shaded region marking a $\pm 1$ standard deviation from this average. Although the load at the initial peak depends on the relative density, as expected, the overall behavior (load peak, subsequent softening, and recovery to a second load peak) was well-established and consistent across specimens of varying relative densities. This averaged response indicated that the qualitative nature of the failure mechanism (i.e., shear banding followed by crushing) is a general feature of the material system.

This sequential XCT strategy allowed us to uncover the extent and nature of internal damage built up in the composite as a function of the deformation. Moreover, a complete three-dimensional reconstruction enabled the visualization of damage throughout the volume of the specimen. In particular, reconstructions indicate that the primary shear band develops along a diagonal plane of the specimen, spanning the thickness of the specimen but localized to a single row of unit cells at each layer (Fig.~\ref{fig:XCT}(b)). Therefore, despite the local maximum in load at the onset of shear band formation, there remains a large portion of pristine unit cells which preserve the load transfer network in the composite, retaining the overall load capacity despite the nucleation of failure. As the deformation progresses, cracks propagate through the carbon phase as well as at the interface between struts and surrounding matrix material. Although there is interfacial failure, the three-dimensional interpenetrating structure of the matrix material prevents complete pull-out- or delamination-like events. 

As macroscopic strain increases, secondary shear bands develop along opposite diagonal planes (Fig.~\ref{fig:XCT}(c)). These diagonal planes all belong to the $\langle110\rangle$ family of crystallographic planes in the specimen coordinate system. Although fracture is widespread at this point, the reconstructions of the transverse plane suggest that the damage at a given cross-section is localized to a ring of unit cells corresponding to shear planes; unit cells outside of this failure process zone remain entirely unaffected. Even as the damage progresses to the crushing stage (Fig.~\ref{fig:XCT}(d)), although the process zone grows in size, a portion of unit cells near the center of the specimen, outside of the shear band regions, remain pristine, and this pattern continues until ultimate failure  characterized by a tortuous, three-dimensional network of cracks through the specimen (Fig.~\ref{fig:XCT}(e)). Together, these observations explain the consistent load capacity and high toughness exhibited by the samples. 

Finally, the sequential XCT reconstructions confirm that cracks in the composite do not originate from the small voids within the continuous matrix phase resulting from the fabrication process (and making up a total of less than 1\% of specimen volume) observed in the specimens prior to compression. Moreover, neither the shear-band regions nor the compaction regions appear to be affected by the existence of these voids. This suggests that the role of voids in the matrix as stress concentrators can be neglected and confirms that the fabrication method for architected composites yields robust materials with a high degree of tolerance for such imperfections.

\subsection{The interaction effect} As depicted in the region shaded in blue in Fig.~\ref{fig:stress_strain}, there is a marked difference in load capacity between the carbon-epoxy composite and the sum of the individual responses of the carbon and epoxy phases. 
This enhanced mechanical performance, which has been noted in the composites literature for a variety of materials and structures \cite{DMello2013, Han2017, Wang2024, White2021}, is typically referred to as the \textit{interaction effect}. In particular, the presence of the matrix phase has enabled a pathway for stress to be distributed throughout the entire height of the specimen, suppresing a layer-wise crushing response and instead leading to failure that spans the entire specimen length. Moreover, the addition of the matrix phase delocalizes failure away from individual nodes which permits the strain corresponding with the peak load to increase. Even after failure begins, it is initially confined to the shear bands, allowing the rest of the composite to sustain the applied stresses without a dramatic loss in load capacity. This stable behavior recalls the similar favorable result of the non-woven composite framework \cite{Das2018} in contrast to the traditional brittle fracture mode of pyrolytic carbon-based structures or the catastrophic failure mode of laminated composites.

We also measured a net positive interaction effect for the polymer-based composite samples which exhibited a similar stable compaction mechanism. In particular, the addition of a matrix material suppressed the foam-like layer-wise compaction behavior seen in the single-material polymer reinforcing phases, which give rise to a sustained plateau stress before densification. Instead, the ability of the matrix to delocalize deformation and transfer load led to macroscopic shear-banding (see Section \ref{sec:polymercomposites}), consistent with other studies on polymer-composite structures \cite{DMello2013, Han2017}, including IPCs \cite{Zhang2021a}. After shear bands formed in the polymer samples, the ductility and incompressibility of the matrix material led to a tearing-dominated failure mode which further increased energy dissipation with applied strain. In section \ref{sec:auxetic}, we describe how we can harness the near-incompressibility of the PDMS matrix to optimize the interaction effect in the polymer-PDMS system.

\subsection{Energy absorption capacity of the composite system} Numerical integration of the composite stress-strain response suggests that the volumetric energy absorption capacity, or toughness, of the carbon-epoxy composites (to a strain of $\varepsilon_f = 50$\%) is 37 $\pm$ 3 MJ~m$^{-3}$. The measured toughness was found to be largely independent of relative density of the reinforcement phase. Using measurements of the as-fabricated density for each sample, the measured range of specific energy absorption (SEA) in our composites, defined to be 
\begin{equation}
    \text{SEA} \equiv \frac{1}{\rho}\int_0^{\varepsilon_f} \sigma(\varepsilon) \; d\varepsilon,
\end{equation}
was 34 $\pm$ 2 J~g$^{-1}$.  The SEA represents the toughness per unit mass of the material, and is a measure of the effectiveness of the IPC to dissipate energy during its deformation.

Typically, architected materials designed for energy absorption attain a high value of toughness due to high ultimate strains, such as in the case of metallic lattices. However, the elevated density of the precursor material renders the SEA for such structures low. On the other hand, carbon-based materials (e.g., carbon nanotubes, carbon foams, and carbon plate-based lattices) can span orders of magnitude in density with a corresponding variation in energy absorption capacity. Polymeric architected materials and polymer-polymer IPCs can also attain high energy absorption capacities, but examples of high-SEA polymer materials are limited.

\begin{figure}
    \centering
    \includegraphics[width=0.98\linewidth]{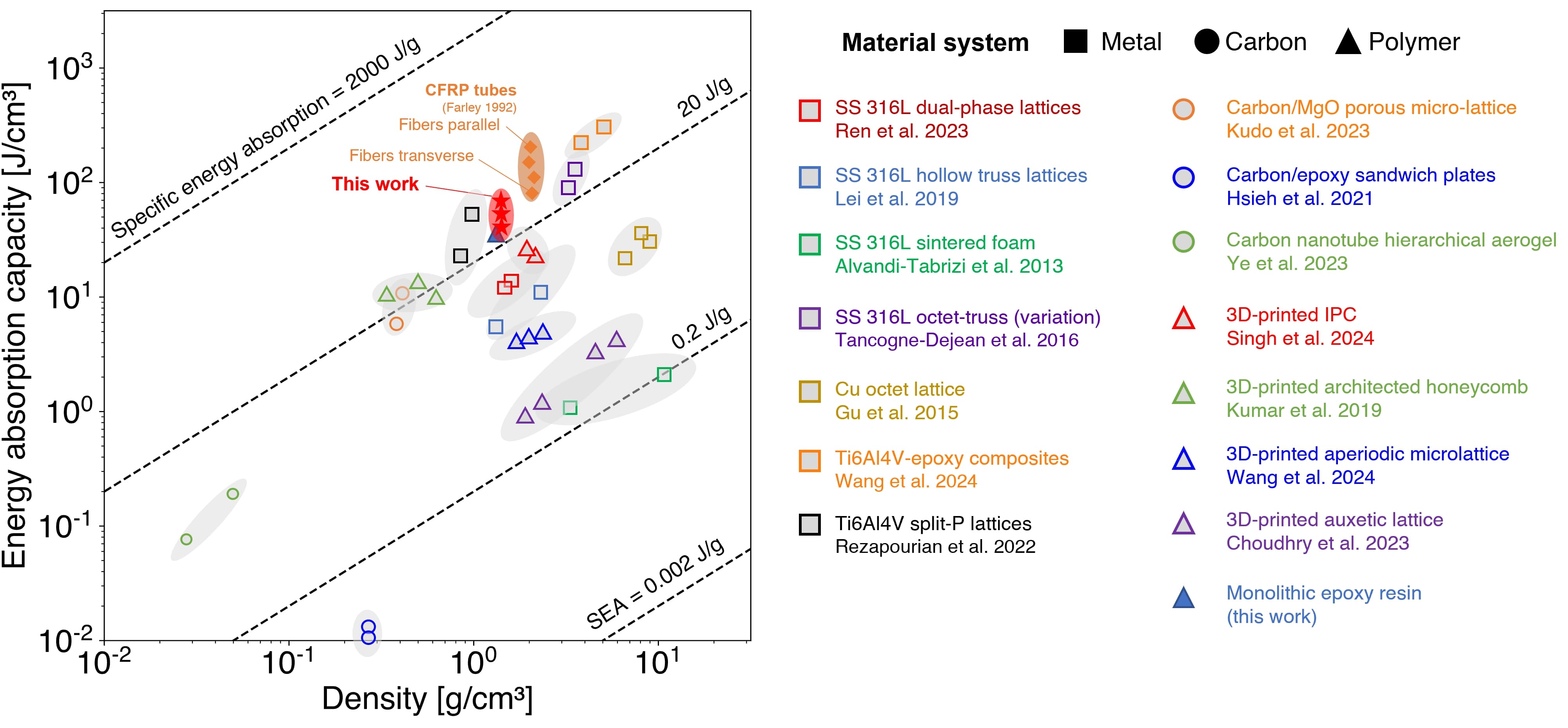}
    \caption{Ashby plot of energy absorption capacity versus material density for single-material and composite architected materials in the current literature \cite{Wang2024, Singh2024, Rezapourian2023, TancogneDejean2019, Ren2023, Lei2019, AlvandiTabrizi2014, WendyGu2015, Kumar2019, Kudo2023, Ye2023, Hsieh2021, XWang2024, Choudhry2023, 10.5555/888240}. Specific energy absorption is represented as lines of constant slope.}
    \label{fig:SEA_ashby}
\end{figure}

Figure \ref{fig:SEA_ashby} is an Ashby plot of energy absorption versus density for a variety of materials and architectures in the literature \cite{Wang2024, Singh2024, Rezapourian2023, TancogneDejean2019, Ren2023, Lei2019, AlvandiTabrizi2014, WendyGu2015, Kumar2019, Kudo2023, Ye2023, Hsieh2021, XWang2024, Choudhry2023}; lines of constant slope represent constant values of SEA. Also plotted is the response of the monolithic epoxy resin used in our experiments, which attains an SEA of 25.8 $\pm$ 0.4 J~g$^{-1}$. Because of its low ultimate strength, the carbon reinforcing phase alone contributes negligibly to toughness compared to the matrix phase alone (cf. Fig.~\ref{fig:stress_strain}); hence, the difference in SEA between the monolithic epoxy resin and the carbon-epoxy composite represents the interaction effect expressed on a per-unit-mass basis. 

On the basis of SEA, carbon-epoxy composites in the present work outperform nearly all examples of architected materials in Fig.~\ref{fig:SEA_ashby}. This is a direct result of the stable compaction mechanism developed in the IPC which permits a sustained, elevated load capacity even after the onset of failure occurs, a phenomenon not seen in single-material architectures, combined with the high inherent strength of pyrolytic carbon not observed in polymer-polymer IPCs.

Furthermore, these composites outperform some carbon fiber reinforced polymer (CFRP) tubes fabricated by fiber winding \cite{10.5555/888240}. Due to the inherent anisotropy of fibers, these wound-fiber tubes are significantly weaker when loaded transverse to the fiber orientation compared to when loaded parallel to the fibers. In this case, our orthotropic carbon-epoxy IPCs fall within the range spanned by wound-fiber tubes, suggesting that this three-dimensional composite framework may find applications where a tunable degree of anisotropy is desired and where the flexibility of freeform fabrication is advantageous, but a high energy absorption capacity is required. 

\section{Architecting a stress state via tunable microstructure}
\label{sec:auxetic}
\begin{figure}
    \centering
    \includegraphics[width=0.9\linewidth]{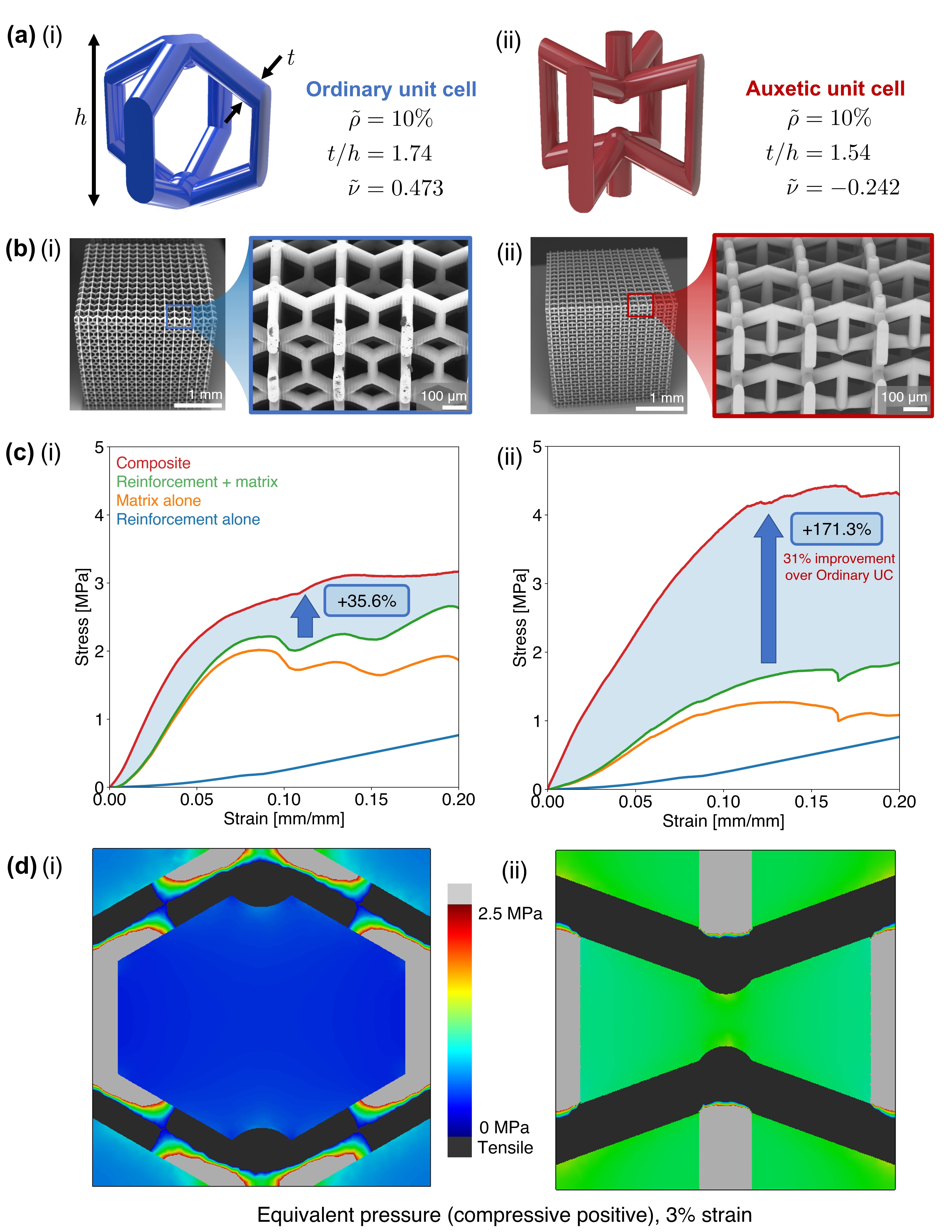}
    \caption{Architecting a stress state in a composite using geometry. Panel (a)(i) shows an ``ordinary'', bending-dominated unit cell with a relative density of 10\%. Panel (a)(ii) shows a unit cell of similar nodal connectivity and equal relative density; however, the geometry of the structure renders the unit cell auxetic, with effective Poisson ratio $\tilde{\nu} = -0.242$. Fabrication by TPL (b) and subsequent infiltration with PDMS shows that the auxetic cell experiences a significant increase in strength and toughness (c), improving in absolute terms even over the composite ordinary cell. Finite element simulations (d) demonstrate that the amount of mean normal pressure is higher in the auxetic unit cell, a consequence of the relation between auxeticity of the structure and near-incompressibility of the matrix material.}
    \label{fig:auxetic}
\end{figure}

Having developed the connection between geometry and behavior in the composite system, we now turn to the idea of architecting not only the morphology, but also the mechanical behavior of an IPC. We consider the two unit cells shown in Fig.~\ref{fig:auxetic}(a), where both architectures have the same relative density ($\tilde{\rho}_f = 10\%$) and hence merely represent a redistribution of mass. Both unit cells have similar nodal connectivities and both fail to satisfy the Maxwell criterion (and are hence bending-dominated). However, while the unit cell in Fig.~\ref{fig:auxetic}(a)(i) has an effective Poisson ratio $\tilde{\nu} = 0.473$, on account of its geometry the unit cell in Fig.~\ref{fig:auxetic}(a)(ii) has $\tilde{\nu} = -0.242$; that is, it displays auxetic behavior whereupon the lateral faces tend to displace inwards when uniaxial compression is applied (Section \ref{sec:auxeticproof}). This unit cell is inspired by the two-dimensional ``bowtie'' geometry that has been well-studied as a classical example of an auxetic morphology \cite{Masters1996, Evans2000}. For simplicity, we henceforth refer to the geometry in Fig.~\ref{fig:auxetic}(a)(i) as the ``ordinary'' unit cell, and to the geometry in Fig.~\ref{fig:auxetic}(a)(ii) as the ``auxetic'' unit cell.

In particular, we consider the case where IPCs of these morphologies are filled with PDMS, which is a material that displays near-incompressible behavior \cite{Huang2005}. Since such materials are capable of sustaining an arbitrary hydrostatic pressure in addition to the stress state associated with a given deformation \cite{Anand2022}, the auxetic unit cell is designed to create a domain in the matrix phase where the equivalent pressure $\bar{p} \equiv -\mathrm{tr}(\boldsymbol{\sigma})/3$ is elevated, and thus contributes to an apparent strength increase in the composite.

To demonstrate this behavior, we fabricated polymer architectures out of UpPhoto (Fig.~\ref{fig:auxetic}(b)) and infiltrated them with PDMS. The polymer system was chosen for this study to ensure uniformity and repeatability of the geometry, as the effective Poisson ratio of the auxetic geometry was found to be sensitive to geometrical parameters. To this end, we tested both uninfiltrated specimens and polymer/PDMS composite specimens in uniaxial compression to determine the effect of adding an incompressible matrix material on mechanical performance.

Figure \ref{fig:auxetic}(c) illustrates the results in the same manner as presented before for the Kelvin geometry: the stress-strain curves of the two single-material phases (architecture alone and matrix alone), their sum, and the stress-strain curve of the composite are each plotted. The interaction effect is again shaded in blue.

In the single-material case, the ordinary geometry outperforms the auxetic geometry in terms of stiffness, toughness, and load capacity. The auxetic unit cell succumbed to a buckling mode due to the presence of aligned, vertically-oriented beams, creating an instability that resulted in a loss of load capacity. In contrast, the ordinary geometry exhibited stable behavior whereby despite local failure at the unit-cell level (marked by local maxima on the stress-strain curve), the remaining pristine unit cells were able to recover (marked by local minima on the stress-strain curve), maintaining a roughly constant plateau stress. In the high-strain limit, both unit cells were compaction-dominated, exhibiting foam-like densification.

In the composite case, the ordinary architecture demonstrated a non-zero but modest interaction effect, gaining a 35.6\% increase in toughness to 20\% strain and a 24.5\% increase in stiffness. The lack of local extrema in the composite stress-strain curve suggests that the matrix material suppresses the effect of local failure by redistributing load appropriately, creating a stable compaction mechanism consistent with that observed for the Kelvin cell. The composite auxetic architecture also experiences stable compaction, but gains a remarkable 171.3\% increase in toughness and a 158.8\% increase in stiffness, and reaches a higher effective plateau stress compared to the composite ordinary architecture. Not only is the matrix in this case able to suppress buckling and local failure, but the combination of geometric auxeticity and incompressibility is sufficient to raise the load capacity of the composite auxetic cell when compared to its ordinary counterpart.

To obtain further insight on the 3D stress state within the composites, we simulated the response of both architecture types with close attention to the pressure within the matrix phase. The matrix phase was modeled as an incompressible Arruda-Boyce polymer using constants $(\mu = 1.337 \, \mathrm{MPa}, \lambda_m = 1.488)$ fit from compression experiments on samples of pure PDMS. The results, shown in Fig.~\ref{fig:auxetic}(d), demonstrate that at the same macroscopic strain, the matrix inside the auxetic unit cell does attain a state of greater equivalent pressure. Under the same boundary conditions, simulation of monolithic PDMS reveals a uniform value $\bar{p} = 0.061$ MPa of equivalent pressure.

Notably, as a consequence of the periodic unit cell assumption, the matrix outside the auxetic unit cell also exhibits an increased hydrostatic pressure, which contributes to the stability of the composite architecture and explains its ability to distribute the applied load. Consequently, despite this material system having a sub-critical stiffness fraction ($E_m / E_f = 1.54 \times 10^{-3}$) the beams in the architecture transition into a pure stretching state. In comparison, although the same assumptions of periodicity were applied to the composite ordinary architecture, the beams in this case exhibit a classical bending response associated with this sub-critical stiffness fraction. This suggests the possibility of another pathway to induce a stretching-dominated response in a bending-dominated morphology. 

Together, these results illustrate that the framework of architected IPCs provides the opportunity for tunablility and optimization of not only morphology but also the ensuing stress state. Moreover, the use of a commonly available, chemically benign and easy-to-handle matrix material like PDMS contributes to the facility of realizing such tunable IPCs. 

\section{Summary}
\label{sec:elastic}
In this work, we demonstrate that the structure of an architected interpenetrating phase composite yields a tunable and robust material system that results in superior mechanical properties. In particular, with the classical scaling law approach for stiffness as inspiration, we identify a critical material parameter of the composite---namely, the ratio of Young's moduli of the constituent phases, $E_m / E_f$---that dictates the ensuing elastic response of the composite. When $E_m / E_f > 10^{-2}$, regardless of the nonlinear properties of the constituent materials, the composite develops a load transfer pathway that induces a global stretching behavior in the beams of the reinforcing phase, regardless of the rigidity of the network.

Moreover, we demonstrated that although pyrolytic carbon behaves in a brittle manner dominated by a fast fracture mode, the addition of a load-bearing matrix suppresses this effect in favor of a stable compaction deformation where the load capacity remains high despite the nucleation and development of shear bands. Consequently, the composite outperforms the sum of its constituent parts in terms of strength and toughness, a measure of the beneficial interaction effect between the two phases. We find that this effect is generalizable to architected composite systems independent of the specific choice of materials, and is rather a realization of effective load transfer under applied stress. Moreover, we show that the specific energy absorption capacity of the carbon-epoxy system is superior to many current examples of architected composites in the literature, due to the unique properties afforded by the strong, lightweight carbon phase together with the resilient epoxy phase. 

More generally, we demonstrate how the interaction effect can be tuned by varying the morphology of the specimen to harness the varying properties of both architecture and constituent materials. We show that when an auxetic unit cell is combined with a nearly-incompressible matrix material, the compaction behavior can switch from an unstable to a stable trend, and the corresponding interaction effect can be optimized to yield a mechanically robust composite.

Altogether, this work represents a step towards fabricating and characterizing scalable three-dimensional architected composites. The carbon-based system presented here represents a promising avenue for optimization, and more work remains to understand the effect of morphology and processing parameters on the resulting composite, as well as how the material may be optimized for a given loading condition. Promising future avenues include the use of shell-based or aperiodic morphologies, which have the potential to remove nodal stress concentrations and harness the beneficial geometric effects of curvature. These and other morphologies are easily realized using the fabrication pathway outlined in this work, broadening the achieveable class of next-generation structural 3D architected composites.

\newpage
\appendix
\section{Supplementary Information}
\label{sec: Appendix}
\subsection{Materials and methods}\label{sec:detailed_methods}
\paragraph{Materials} For TPL 3D-printing, acrylate-based two-photon photosensitive resin UpPhoto was used as received (UpNano, Vienna, Austria). For VP 3D-printing, poly(ethylene glycol) diacrylate (PEGDA, average molecular weight 700), photoinhibitor Sudan I, and photoinitiator phenylbis(2,4,6-trimethylbenzoyl)phosphine oxide (BAPO) were used as received (Sigma-Aldrich, USA). For matrix infiltration, silicone elastomer Sylgard 184 (PDMS, Dow Corning, USA), urethane resin Smooth-Cast 61D (Smooth-On, USA), and two-part epoxy resin West System 105/206 (West System, USA) were used as received.

\paragraph{Two-photon lithography} Structures were printed using a pulsed femtosecond laser having center wavelength 780 nm focused through a 10$\times$ optical objective. The laser operated at 100 mW, scanning at 600 mm s$^{-1}$, and the hatching and slicing distances were set to 2.1 \textmu{}m and 3.0 \textmu{}m, respectively. In the NanoOne software suite Think3D, each 4 mm specimen was subdivided into 60-\textmu{}m tall ``blocks'' for printing, with a 50\% block offset between neighboring blocks. The block offset procedure ensures that the printed structure does not delaminate along stitching lines defined by the objective field-of-view. No such delamination was observed during subsequent experiments. Structures were printed onto a cleaned and silanized glass substrate. Following the manufacturer's recommendations, printed parts were washed in 2-propanol (Sigma-Aldrich, USA) for a minimum of 30 minutes, followed by drying for a minimum of 24 hours in air. No further post-processing of printed parts was performed. After drying, the printed parts were carefully removed from the substrate using a razor blade.

\paragraph{Vat photopolymerization 3D-printing} We formulated a custom, poly(ethylene glycol) diacraylate (PEGDA)-based resin for VP printing by combining PEDGA, BAPO, and Sudan I at a ratio of 100:1:0.1 by mass. The resin was ultrasonicated for 60 minutes at room temperature and used thereafter for printing. Exposure times for 25 \textmu{}m-tall layers were 18 s for the first ten layers, followed by 8 s for all layers thereafter. After printing, the PEGDA structures were washed in 2-propanol for a minimum of 12 hours, followed by drying for a minimum of 24 hours in air.

\newpage

\subsection{Constituent material properties}\label{sec:matprops}
The mechanical properties of each constituent material were measured by performing uniaxial compression experiments on monolithic samples at the same loading conditions as all other experiments, with the exception of pyrolytic carbon, where Berkovich nanoindentation was used to obtain a value of the reduced stiffness \cite{Oliver1992}. We performed a minimum of three replicate experiments on each material, and a representative stress-strain curve for each material is shown here.

\paragraph{Reinforcement materials} Using the particular printing parameters discussed above, polymerized UpPhoto is an elastic-plastic material with post-yield hardening. The elastic modulus was measured to be $2.5 \pm 0.2$ GPa and the yield strength was measured to be $92 \pm 8$ GPa. The elastic modulus of the pyrolytic carbon was measured to be $32 \pm 4$ GPa, in agreement with values for disordered pyrolytic carbon found in the literature \cite{Eggeler2023}.

\paragraph{Matrix materials} PDMS exhibited a stress-strain response in uniaxial compression consistent with that of an incompressible, nonlinear hyperelastic material with the Arruda-Boyce parameters given in the text. In the limit of small strains, the measured elastic modulus was $4.1 \pm 0.3$ MPa. Urethane resin Smooth-Cast 61D exhibited a nonlinear response with elastic modulus $80 \pm 4$ MPa, but exhibited permanent deformation upon unloading. Epoxy resin was measured to have an elastic modulus of $845 \pm 69$ MPa with a clear elastic-plastic response consistent with that of a highly-crosslinked polymer \cite{Ashby2013}.

\begin{figure}[h!]
    \centering
    \includegraphics[width=0.8\linewidth]{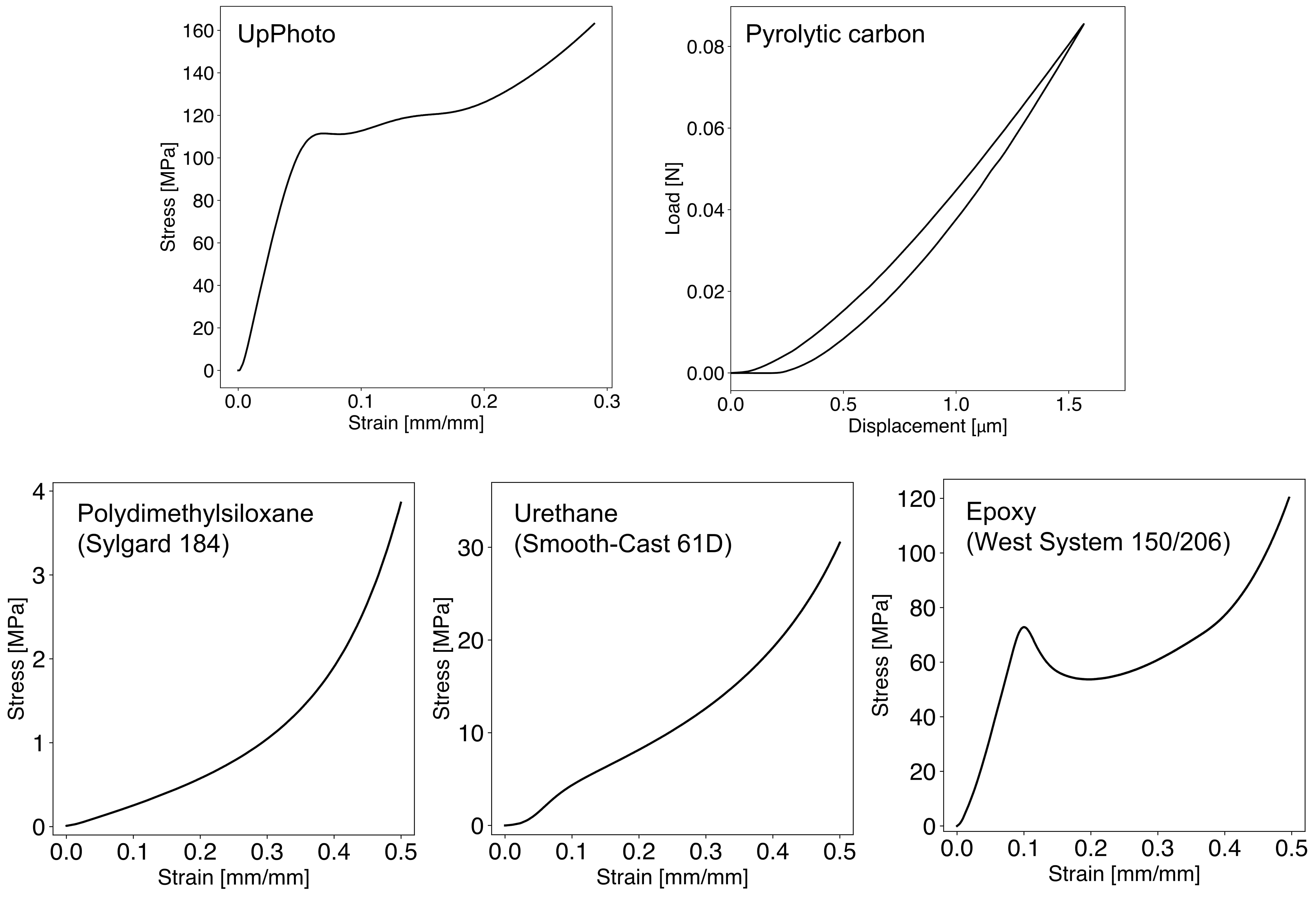}
    \caption{Typical mechanical response in uniaxial compression (or indentation, in the case of pyrolytic carbon) for the constituent materials used.}
    \label{fig:SI-constituent}
\end{figure}

\newpage

\subsection{Characterization of architected IPCs} \label{sec:detailed_characterization}
\paragraph{Microscopy} The fabricated composites were visually inspected after each fabrication step using optical and scanning electron microscopy. Optical microscopy was used to verify the morphology of features but was not used to make critical measurements. Using a desktop scanning electron microscope (SEM; Hitachi, Tokyo, Japan), micrographs were taken of each specimen at a minimum of three distinct locations across its exterior faces (Section \ref{sec:sem}). Using these micrographs, a commercial image analysis software (ImageJ, National Institutes of Health/LOCI, University of Washington) was used to measure beam thicknesses and unit cell sizes. An average of at least three such measurements was used to compute the as-fabricated relative density for each sample.

\paragraph{X-ray computed tomography} X-ray computed tomography (XCT) scans of the fabricated composites was performed using a Versa 620 X-ray microscope (Zeiss, Germany). Samples were mounted on a thin graphite sample holder. X-rays were generated at 60 kV with a total power of 6.50 W, and 4-second exposure scans were taken using the 0.4$\times$ objective through air. A typical voxel size was 7.65 \textmu{}m. To perform three-dimensional reconstruction of sample morphology, 1601 scans were taken as the sample rotated through one full revolution. Post-processing and reconstruction were performed with a commercially available software suite from the manufacturer and with ImageJ.

\paragraph{Mechanical testing} Uniaxial compression experiments was performed on an Instron 5900 series universal testing system (Instron, USA) under displacement control at a strain rate not exceeding $10^{-3}$ s$^{-1}$. The polymer-based system was measured using a 1 kN load cell to maximize sensitivity and the carbon-based system was measured using a 50 kN load cell due to higher ultimate loads. Prior to the application of deformation, macroscopic specimen dimensions were measured; these dimensions were used to convert load and displacement data recorded by the universal testing system to stress and strain, respectively.

 In accordance with the ASTM standard protocol for uniaxial compression (D695) \cite{ASTM_D695}, the elastic modulus of each sample was extracted using a least-squares regression linear fit on the initial portion of the stress-strain curve, excluding any initial nonlinear ``toe'' region corresponding to system compliance. The yield strength of each sample was extracted using the standard 0.2\%-offset method, and the toughness of each sample was computed by numerical integration of the stress-strain data.

\paragraph{Computational methods} To numerically simulate the behavior of architected IPCs, we modeled a single unit cell as a two-phase structure consisting of a reinforcing phase and a matrix phase, each having independent constituent material properties derived from experiments. Geometries were exported from CAD software as solid-body files and imported into the commercial software ABAQUS, where they were joined using a Boolean operation which preserved geometrical boundaries and individual material assignments. The combined body was meshed using second-order tetrahedral elements (C3D10) unless the material was modeled as incompressible (e.g., in the case of PDMS), for which second-order tetrahedral elements with hybrid formulation (C3D10H) were used instead. The ideal mesh seed size was determined as a result of a mesh convergence study  to be a linear function of the beam thickness $t$. Namely, the convergent seed size was taken to be $t \times (0.3 / 1.1)$, which created four mesh elements across the diameter of each beam, independent of beam diameter, typically yielding 80,000 to 120,000 elements for a composite Kelvin cell.

To determine the optimal mesh size for FE simulations, the mesh was optimized based on the figure of merit $E_c / E_f$ for a given composite unit cell with fixed $E_m$ and $E_f$. Linear perturbation studies to 1\% macroscopic strain were performed on this unit cell with periodic boundary conditions. To ensure that adequate spatial resolution was realized for all relative densities, the mesh parameter was taken to be a function of beam thickness $t$. The optimal mesh seed size was found to be $t \times (0.3/1.1)$; this value gave a convergent value of the figure of merit with the least computation time.

\begin{figure}[h!]
    \centering
    \includegraphics[width=0.9\linewidth]{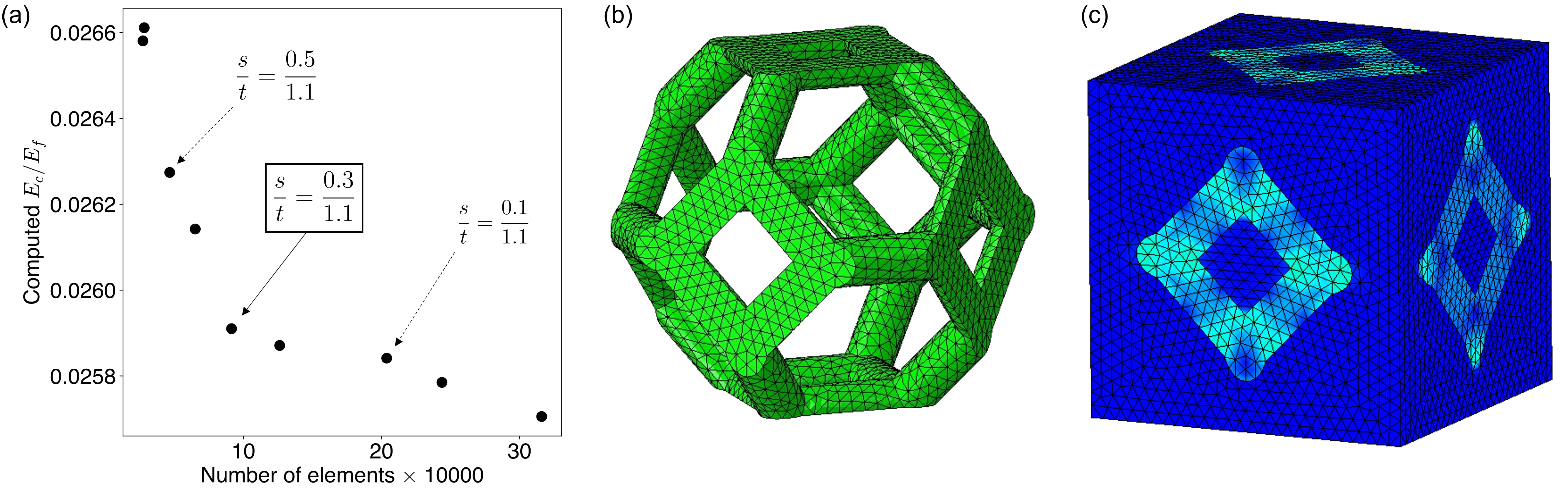}
    \caption{Mesh convergence study, demonstrating that a ratio of mesh seed size $s$ to beam thickness $t$ produces a convergent value of the composite stiffness $E_c / E_f$ (for fixed $E_f$ and $E_m$) at minimum computation time (a). The resulting mesh of the architecture (b) and composite (c) showing a typical element size with four elements across the beam thickness.}
    \label{fig:SI-mesh_convergence}
\end{figure}

To simulate an infinite tessellation of unit cells, we applied a periodic mesh and enforced periodic conditions (PBCs) on the unit-cell model across pairs of matching faces. In addition to these boundary conditions, displacement control was applied using the virtual node approach \cite{Danielsson2002, Okereke2013} to investigate the subsequent deformation behavior and stress distribution within the two phases.

\newpage

\subsection{Pyrolysis parameters}\label{sec:pyrolysis}
\begin{figure}[h!]
    \centering
    \includegraphics[width=0.95\linewidth]{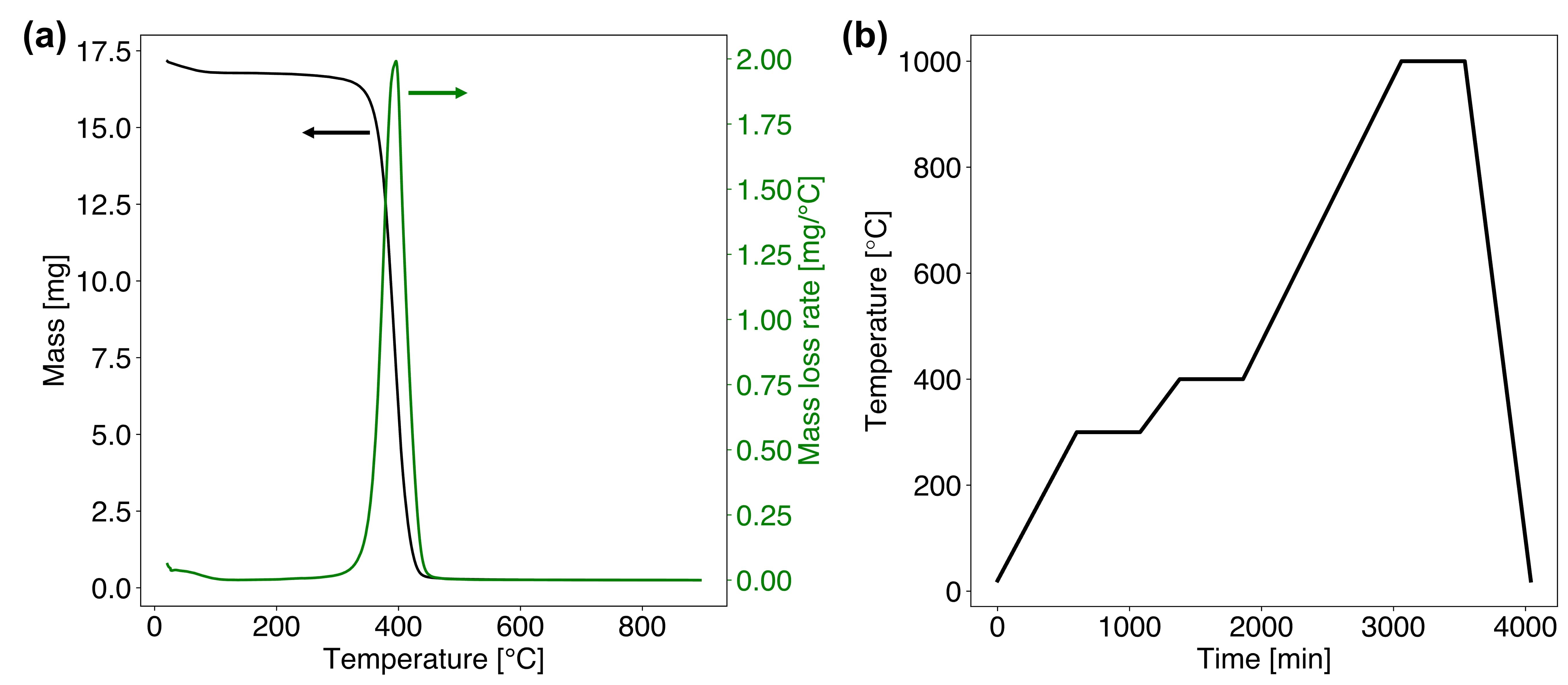}
    \caption{(a) Thermogravimetric analysis and (b) resulting pyrolysis profile for the PEGDA-based custom resin used for carbon conversion.}
    \label{fig:SI-tga}
\end{figure}

Thermogravimetric analysis (TGA) was performed on a sample of the PEGDA-based polymer to determine the pyrolysis profile. We prescribed a heating rate of 5$\degree$C min$^{-1}$ to a final temperature of 900$\degree$C. To replicate the inert conditions of the vacuum furnace, nitrogen gas (N$_2$) was flowed continuously at a flowrate of 100 mL min$^{-1}$ during the process. Figure \ref{fig:SI-tga}(a) shows the mass loss of the sample as a function of temperature.

Based on the TGA data, we prescribed the heating profile in Figure \ref{fig:SI-tga}(b) during the pyrolysis cycle. The extended holds correspond to the polymer degradation regime, and correspondingly the slow ramp rate of temperature ensured uniformity in shrinkage. After the degradation process, the furnace is brought to its maximum pyrolysis temperature of 1000$\degree$C as described in the main text.
\newpage

\subsection{Raman spectroscopy on the pyrolytic carbon}\label{sec:raman}
\begin{figure}[h!]
    \centering
    \includegraphics[width=0.6\linewidth]{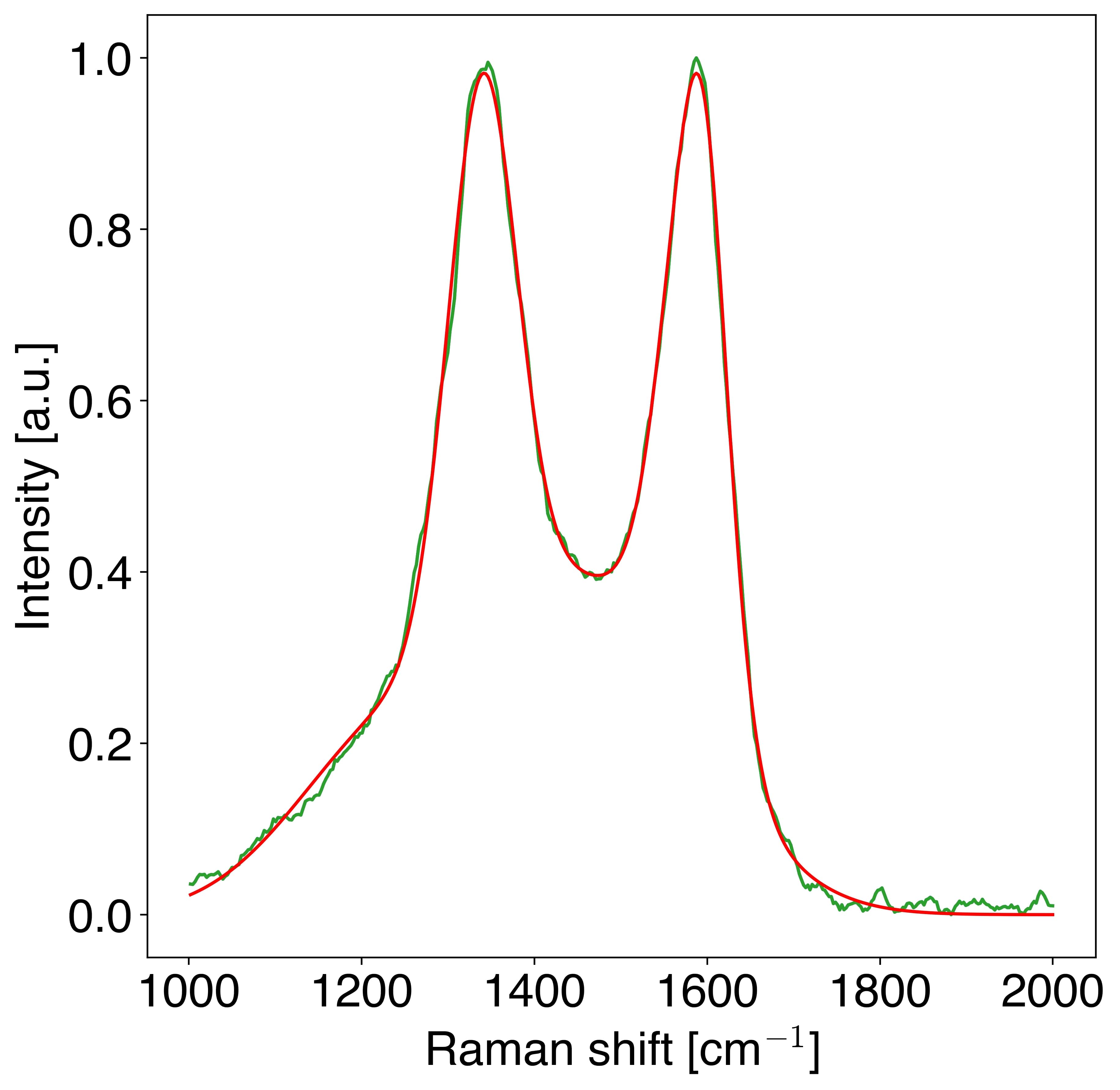}
    \caption{Typical Raman spectrum (green) of pyrolytic carbon lattices and best-fit deconvolution (red) of peaks into those due to ordered and disordered phases.}
    \label{fig:SI-raman}
\end{figure}

We performed Raman spectroscopy on the carbon samples following pyrolysis to determine the degree of graphitization. Measurements were taken using an a Witec alpha300 apyron confocal Raman microscope with a laser operating at 30 mW with an excitation wavelength of 532 nm. The laser spot was focused through a 10$\times$ optical objective, and ten, 2-second accumulations were averaged to obtain a signal. We deconvolved the raw data and fit the known peaks corresponding to vibration modes of the ideal graphite ($G$), disordered graphite ($D1$, $D2$, $D4$), and amorphous carbon ($D3$) phases \cite{Sadezky2005}. The relative degree of graphitization $R_2$, defined to be 
$$R_2 \equiv \frac{I_\text{D1}}{I_\text{G} + I_\text{D1} + I_\text{D2}},$$ where $I_x$ represents the area of peak $x$, was found to be $0.51 \pm 0.03$ across nine carbon samples. A typical Raman spectrum (green) and the sum of the best-fit deconvolution (red) is shown in Figure \ref{fig:SI-raman}.

\newpage

\subsection{Void fraction measurement}\label{sec:voids}
Measurement of the void volume in the composite after infiltration was performed using image processing on individual XCT reconstruction slices. Slices generated by the reconstruction software were exported as image files, and thresholding was performed on each image individually to determine the area fraction of each slice corresponding with air (i.e., a void). These results were then aggregated to compute the total void volume of the specimen. The average void fraction was $0.64 \pm 0.11$\%.

\begin{figure}[h!]  
    \centering
    \includegraphics[width=0.95\linewidth]{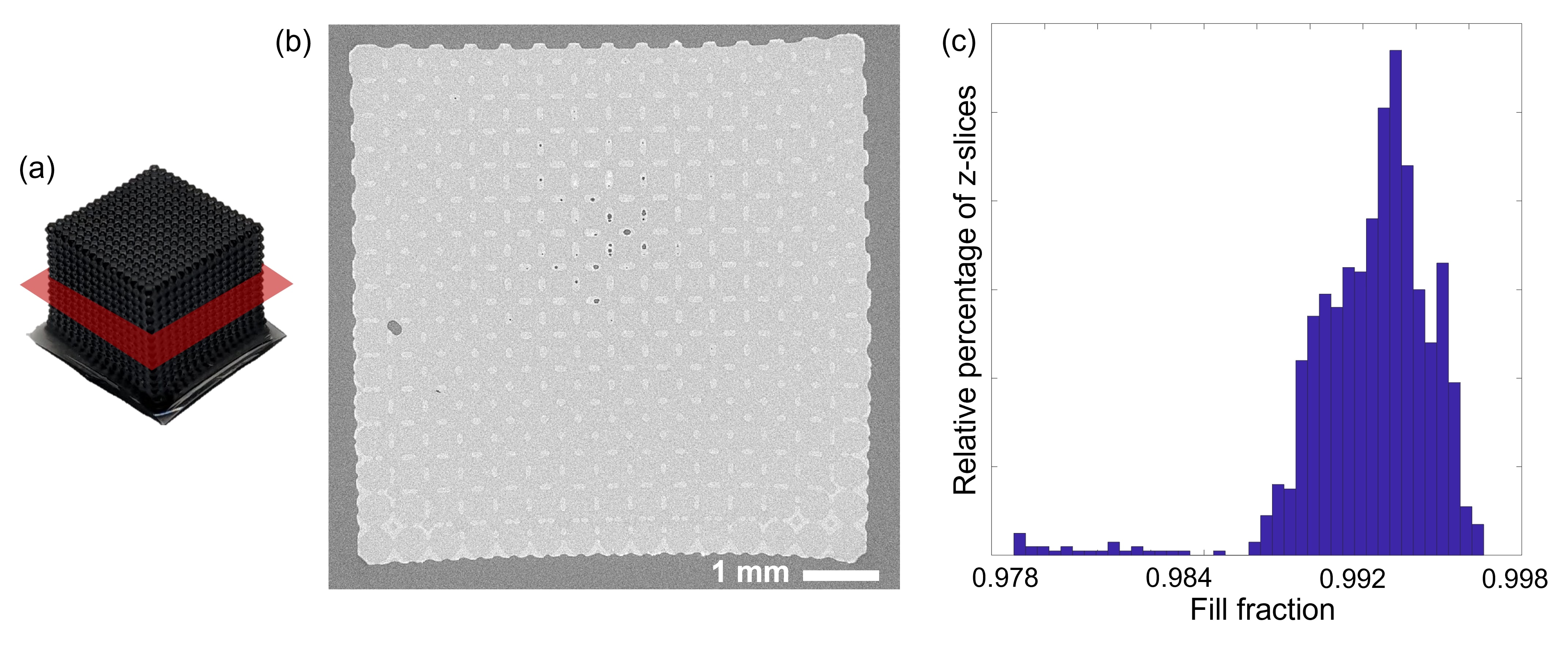}
    \caption{Void volume measurement. (a) Schematic of the representative cross-section. (b) Typical cross-section XCT image (``Z-slice'') before post-processing. (c) Typical histogram over all such Z-slices after image post-processing, showing that nearly all Z-slices had a fill fraction exceeding 99\%. The total void volume of all specimens used for mechanical experiments in this study had a fill fraction $>99$\%.}
    \label{fig:SI-voids}
\end{figure}
\newpage

\subsection{Typical SEM images}\label{sec:sem}
\begin{figure}[h!]
    \centering
    \includegraphics[width=0.9\linewidth]{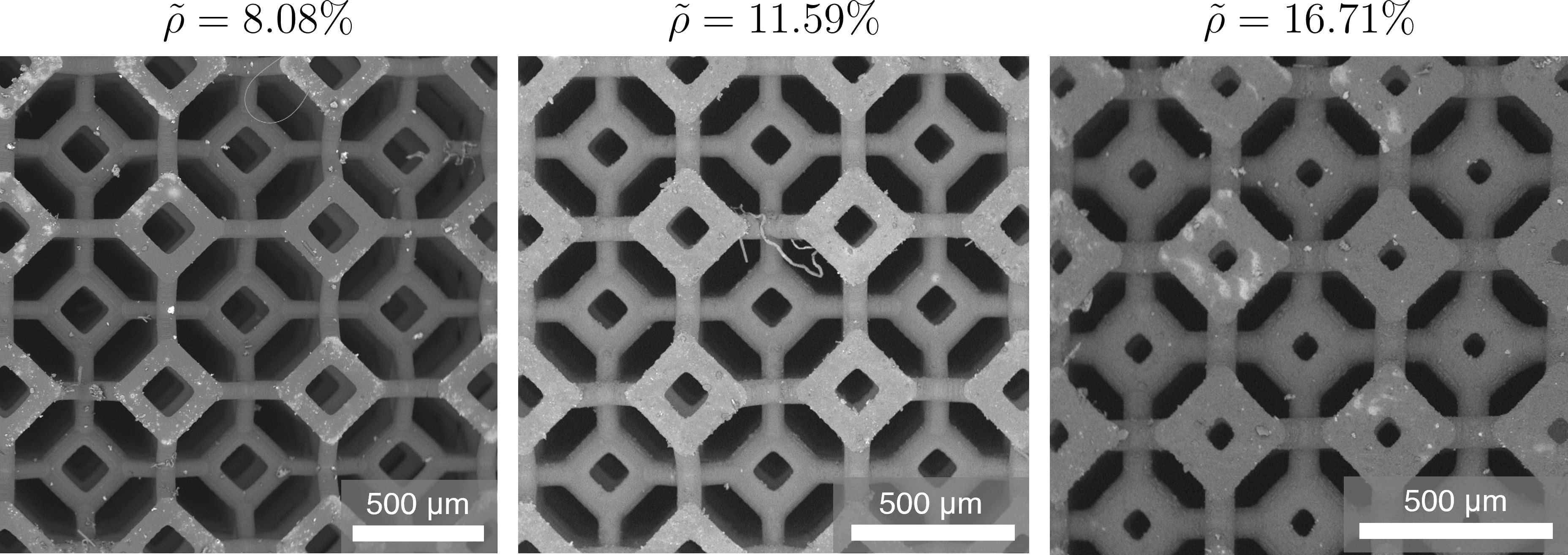}
    \caption{Typical scanning electron micrographs of pyrolytic carbon lattices.}
    \label{fig:SI-sem}
\end{figure}
\newpage

\subsection{Relative-density-dependent behavior of the carbon-epoxy system}\label{sec:rd}
\begin{figure}[h!]
    \centering
    \includegraphics[width=0.8\linewidth]{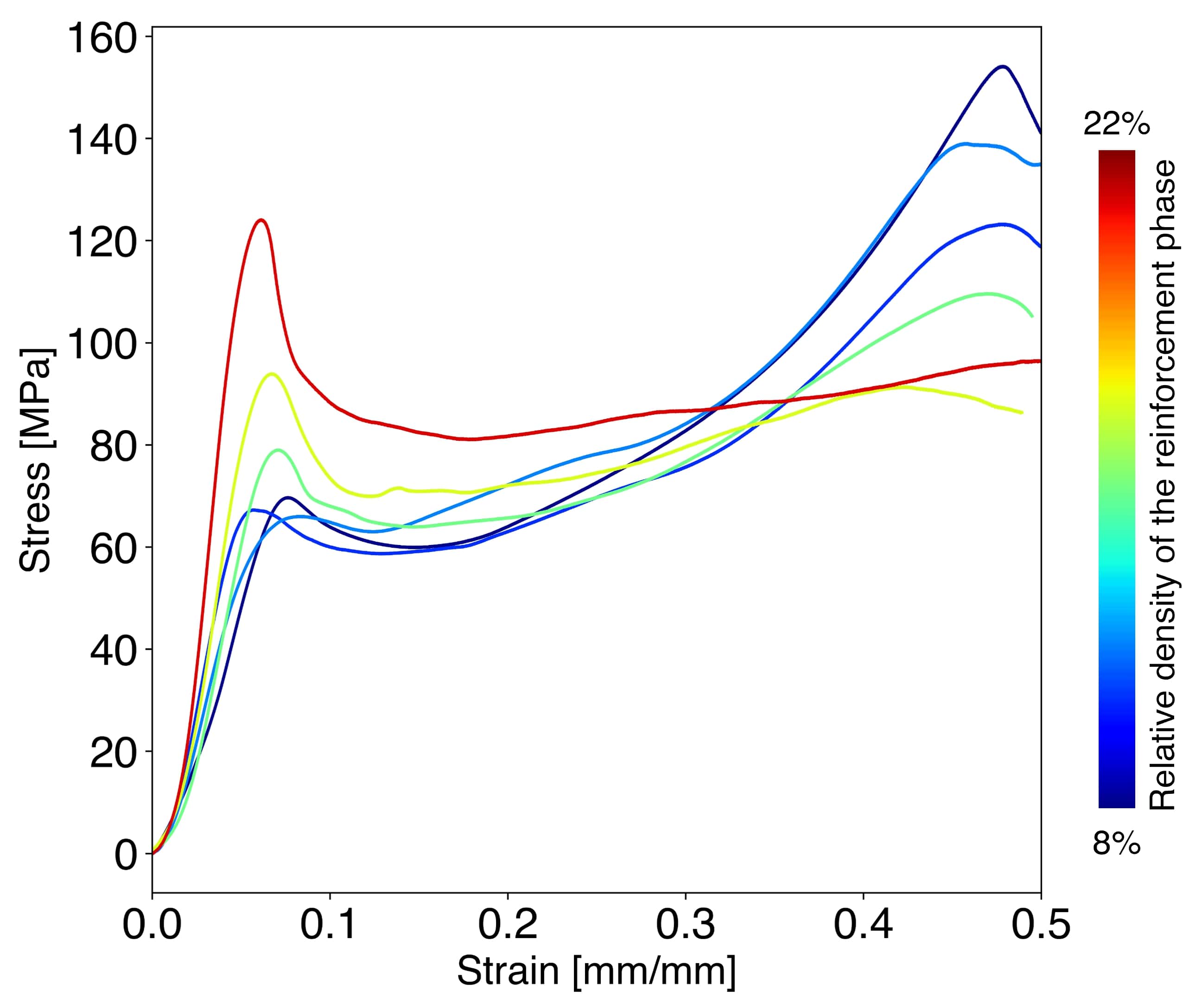}
    \caption{Typical stress-strain curves of carbon-epoxy composites.}
    \label{fig:SI-stress_strain}
\end{figure}

To illustrate the effect of changing the relative density of the carbon reinforcing phase on the composite system, Figure \ref{fig:SI-stress_strain} plots six stress-strain curves corresponding to samples ranging from $\tilde{\rho} = 8$\% to $\tilde{\rho} = 22$\%. In general, the stiffness and strength (i.e., initial peak stress) increase with increasing $\tilde{\rho}$, but because the subsequent post-peak hardening becomes less pronounced with increasing $\tilde{\rho}$, the toughness of the composite samples did not demonstrate a trend as $\tilde{\rho}$ was varied.

\newpage

\subsection{Details on the mechanical performance of the polymer system}\label{sec:polymercomposites}
\paragraph{Single-material behavior} The unfilled polymer lattices exhibited relative-density dependent behavior ranging from foam-like compaction at low relative densities to fracture at high relative densities. For $\tilde{\rho} \lesssim 12\%$ the slender structures tended towards deformation characterized by single-layer compaction bands which spanned the width of the sample. The corresponding stress-strain behavior past the initial linear regime was characterized by a sustained plateau stress until the onset of densification, at which point the stress increased without bound. For $\tilde{\rho} \gtrsim 22\%$ the polymer samples exhibited only a short plateau until a fast fracture event. Fracture typically occurred along a shear band spanning one major diagonal of the structure, and resulted in total loss of load capacity in the specimen. In the intermediate regime, $12\% \lesssim \tilde{\rho} \lesssim 22\%$, features of both the compaction mode and the fracture mode were present. Namely, specimens in this relative density regime experienced both layer-wise compaction and ultimate fracture, and attained a peak stress that was intermediate to the low- and high-relative-density regimes indicated. Post-mortem microscopy revealed that in all cases, micro-cracks in the vertical direction originated from the diamond-shaped features of the Kelvin unit cell. Altogether, these observations agree well with established results for the deformation and fracture behavior of Kelvin foams \cite{Gong2005, Kucherov2014}.

\paragraph{Polymer composite behavior}
We fabricated two composite systems using the polymeric reinforcing phase: a polymer/PDMS system and a polymer/urethane system. In all experiments with either material, the plateau behavior was entirely eliminated in favor of a stable compaction behavior characterized by increasing stress (i.e., an increase in load capacity) with increasing applied strain. This suggests that as in the carbon-composite system, the presence of a load-bearing matrix is sufficient to redistribute the load in a way that delocalizes compaction-like deformation. This observation is in good agreement with existing data on polymer-polymer composites \cite{Carlsson2024_exp}. 

In the polymer systems, large-scale cracks developed progressively (cf. the catastrophic failure mode of the polymer single-material samples) and were accompanied by matrix tearing together with tortuous cracks throughout the reinforcing phase. No delamination- or pull-out-like events were observed; rather, the beams of the reinforcing phase were held in place by the surrounding matrix, despite the progression of failure. In the low and intermediate relative-density regimes of the polymer/PDMS system, widespread local cracking (at the unit cell level) was observed together with global tearing (at the specimen level). In these specimens, global failure followed the shear band planes of the specimen but was not entirely localized there. At the high relative-density regime of the polymer/PDMS system, local and global cracking was constrained to a few unit cells around a global shear band. This gave rise to a sudden failure mode characterized by a sudden load drop. In contrast, for all all polymer/urethane composites regardless of relative density, local cracking of unit cells dominated, and these cracks coalesced in vertical planes as macroscopic deformation was increased. This led to a stable compaction behavior with no sudden failure.

Hence, we hypothesize two mechanisms for obtaining a net positive interaction effect in the polymer-polymer composite systems. First, the redistribution of load by the matrix phase contributes to increased strength, and the tortuosity of the induced crack-path combined with the tearing mode of the matrix contributes to increased energy dissipation over the course of the gradual failure. Second, for the composites which exhibit stable compaction (\textit{viz}. the low and intermediate relative-density polymer/PDMS composites and all polymer/urethane composites), the delocalization of global fracture recruits more unit cells in taking a critical amount of load.

\begin{figure}[h!]
    \centering
    \includegraphics[width=\linewidth]{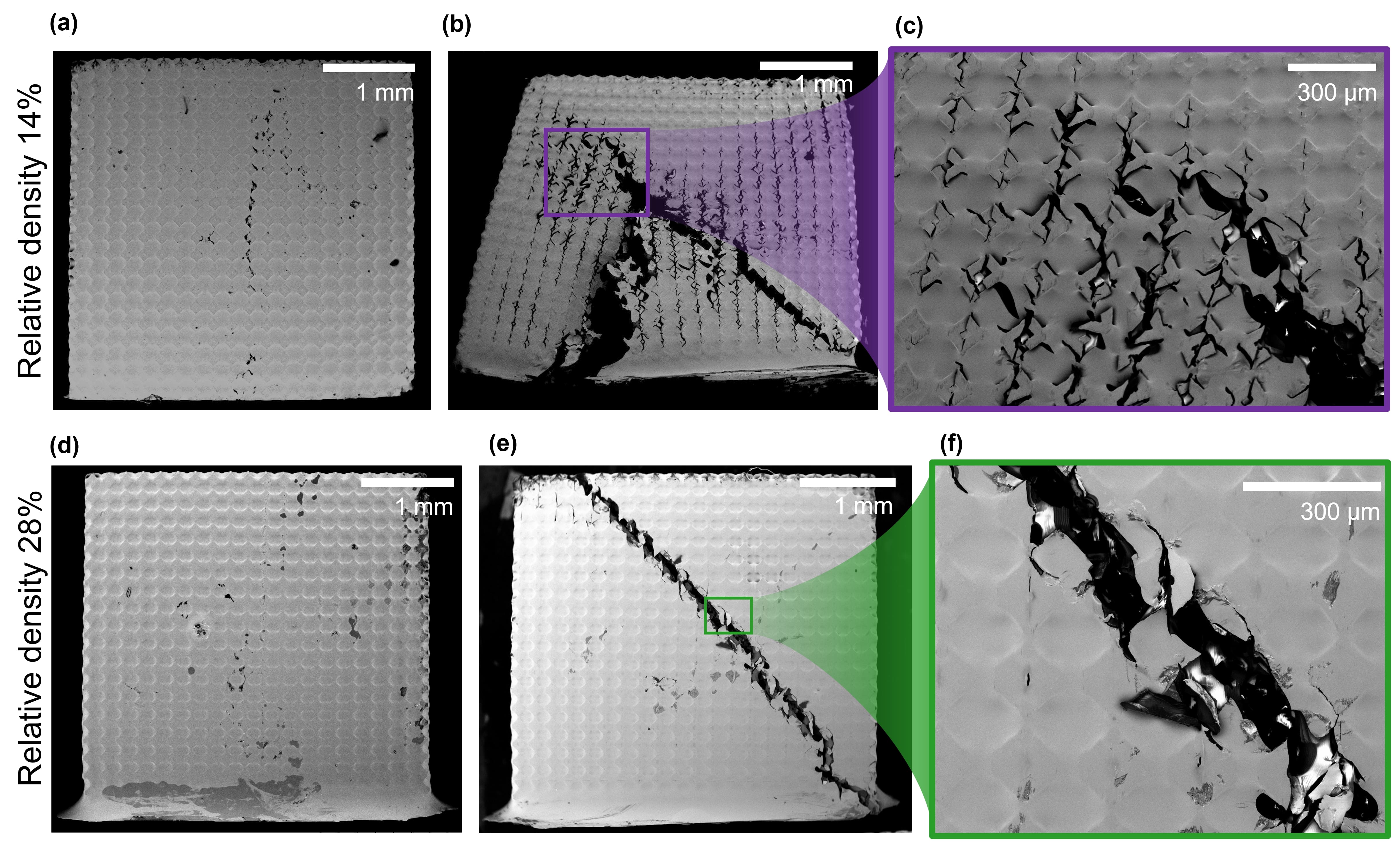}
    \caption{Representative SEM images of polymer/PDMS composites before and after uniaxial compression. In the low-to-intermediate relative density regime, cracking was observed simultaneously with tearing. In contrast, high relative density samples exhibited only cracking along localized shear planes.}
    \label{fig:SI-polymer_composite}
\end{figure}

\newpage

\subsection{Auxeticity of the auxetic lattice}\label{sec:auxeticproof}
\begin{figure}[h!]
    \centering
    \includegraphics[width=0.9\linewidth]{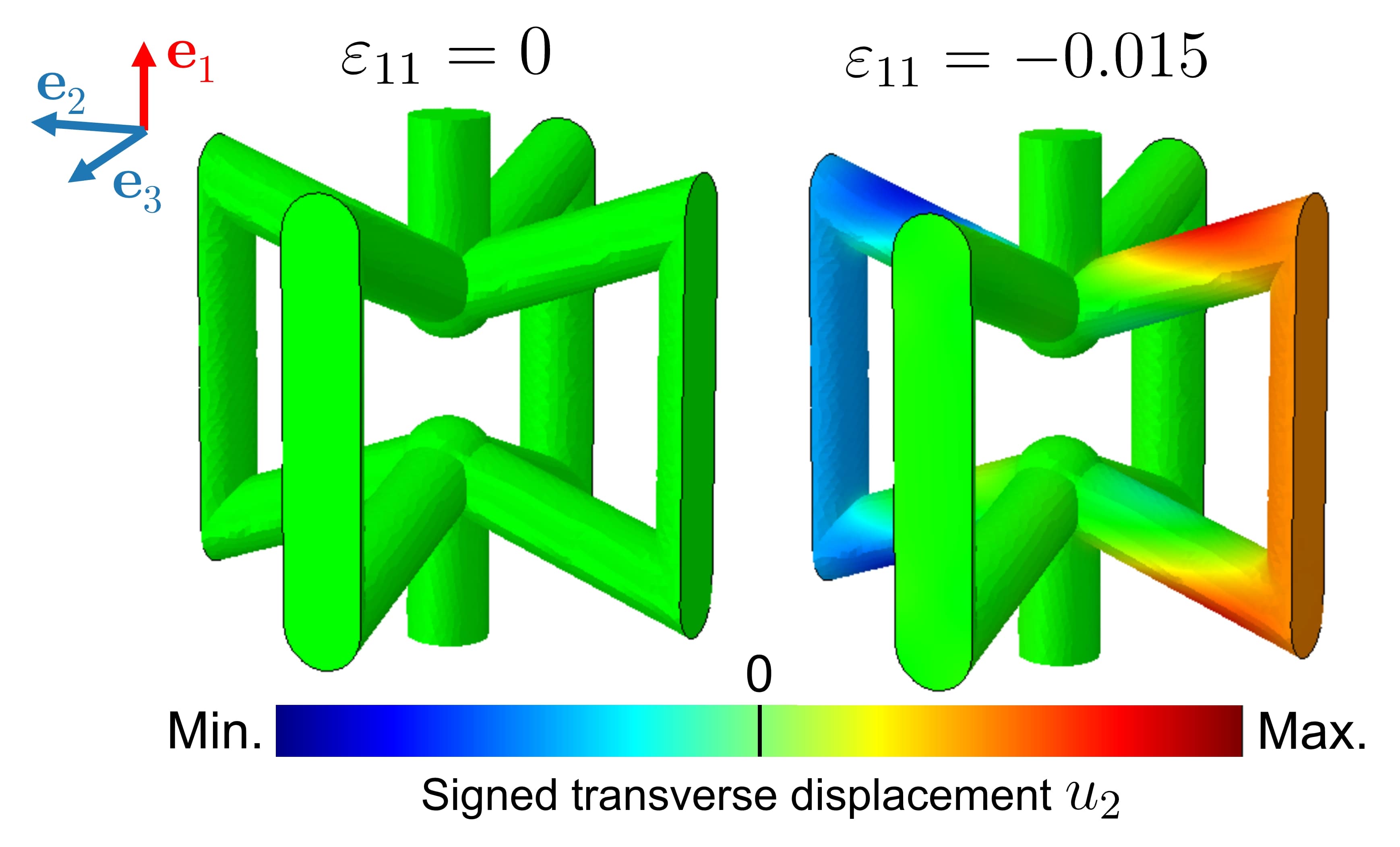}
    \caption{Signed transverse displacement $u_2$ of the auxetic unit cell to a small compressive perturbation in the indicated $\mathbf{e}_1$ direction.}
    \label{fig:SI-auxetic}
\end{figure}

To demonstrate the behavior of the auxetic lattice, we simulated its behavior in the single-material case to a small compressive vertical strain of 1.5\%. We used the experimentally-determined properties of UpPhoto with periodic boundary conditions. As shown in Figure \ref{fig:SI-auxetic}, the lateral sides tend to displace towards the center of the unit cell. The effective Poisson ratio of this geometry was calculated by measuring this lateral displacement relative to the applied vertical displacement.

\newpage
\bibliographystyle{elsarticle-num} 
\bibliography{main_arXiv}

\end{document}